\documentclass[epjCONF]{svjour}
\usepackage{graphicx}
\usepackage{amsmath}
\usepackage{amssymb}
\usepackage{bm}
\usepackage[varg]{txfonts} 
\usepackage[latin1]{inputenc}
\newcommand{\beq}{\begin{equation}}
\newcommand{\eeq}{\end{equation}}
\newcommand{\beqa}{\begin{eqnarray}}
\newcommand{\eeqa}{\end{eqnarray}}
\newcommand{\bs}[1]{\ensuremath{\boldsymbol{#1}}}
\newcommand{\gcmiq}{\, \text{g} \, \text{cm}^{-3}}
\newcommand{\kf}{k_{\rm F}}
\newcommand{\fmi}{\, \text{fm}^{-1}}
\newcommand{\vlowk}{V_{{\rm low}\,k}}
\newcommand{\qq}{{\bf q}}
\newcommand{\om}{\omega}
\newcommand{\kk}{{\bf k}}
\newcommand{\pp}{{\bf p}}
\session-title{%
19$^{\textnormal{\footnotesize th}}$ International %
IUPAP Conference on Few-Body Problems in Physics%
}
\begin{document}
\title{%
Electroweak reactions with light nuclei
}%
\author{%
S. Bacca\inst{1}\fnmsep\thanks{\email{bacca@triumf.ca}} 
}
\institute{%
$^{a}$TRIUMF, 4004 Wesbrook Mall,
Vancouver, B.C. V6J 2A3, Canada
}
\abstract{
The investigation of light nuclei with ab-initio methods 
provides an optimal setting to probe our  knowledge on nuclear forces,
because the few-nucleon problem can be solved accurately.
Nucleons interact not only in pairs but also via many-body forces.
Theoretical efforts  need to be taken towards the 
identification of nuclear observables sensitive to the
less known many-nucleon forces.
Electromagnetic reactions can potentially provide useful information 
on this.
We present results on photo-absorption and electron scattering off
light nuclei, 
emphasizing  the role of three-body forces and the comparison with experimental data.
On the other hand, reactions induced by weak probes, like neutrino interactions with nucleonic matter, are relevant
to astrophysics and can be calculated with few-body techniques. In this case, since
often no experiment is possible,  ab-initio predictions provide valuable input for
astrophysical simulations.  } 
\maketitle
%
%
%
\section{Introduction}
\label{intro}
Excitations of nuclei with external electroweak probes are a fundamental tool
to get information about nuclear dynamics. 
If the probe is energetic enough, it can excite the system such that disintegration
channels are open, and nucleons or compound fragments are emitted. In this case one would talk about an electroweak induced reaction.
This is almost always the case if one considers light nuclei, as they possess no or
very few bound excited states. In general, since a reaction implies states belonging to the
continuum spectrum of the nucleus,  exact direct calculations are difficult and become prohibitive with increasing mass number $A$.  One is then forced to introduce some approximations,
that can be possibly validated against experiment. Such an approach is however not always possible.
The first reason is that in nuclear physics the way the nucleons interact with each other, described by the Hamiltonian, is
the object under investigation. Thus, comparing an approximate calculation with experiments
can pollute a clear interpretation on the reliability of the Hamiltonian used. 
A second reason is that often reactions that are relevant to astrophysics cannot be tested
 experimentally. That is why exact ab-initio calculations
are needed. With this terminology we denote  calculations where all nucleons
are treated as degrees of freedom (ab-initio approach) and where the quantum many-body problem is solved  numerically without  introducing approximations (exact).  Such calculations can be done for light nuclei. They
  can provide a valuable test on our present knowledge of nuclear forces via a comparison with experimental data, and can also give useful predictions to be used e.g. in astrophysical simulations,
when experiments are not possible.  

The interactions among nucleons are governed by quantum cromodynamics (QCD). In the low energy
regime relevant to nuclear structure and reactions, QCD is not perturbative and thus difficult
to solve. Nuclear theory has come up with a series of models to describe the interaction
among nucleons in terms of  {\it effective} degrees of freedom,  protons and neutrons.
Several approaches have been investigated, starting from purely 
 phenomenological ones, or using meson exchange theories and more recently via the 
 systematic approach of effective field theory \cite{EM,EGM}. 
In fact various  
realistic two-body potentials  are available, that  fit  
 N-N scattering data with high accuracy.
On the other hand, since the nuclear potential has clearly an {\it effective} nature, it is in principle a many-body operator. 
Furthermore, it is well known that most realistic potentials do not explain 
the binding energy of three-body nuclei, unless three-body forces  (3NF) are introduced.
Today, aided by effective field theory, where three-nucleon forces arise naturally
and consistently with two-nucleon forces, a new debate is taking place regarding the role of 3NFs and how to constrain them.
For the determination of a  {\it realistic}  three-body potential or 
to discriminate among different models one needs to find  A$\geq3$ observables that are sensitive to 3NFs. 
An important activity in this direction has taken place in recent years, 
with accurate calculations of bound-state properties of nuclei of increasing mass number A
\cite{Pieper02,Navratil07}. 
A complementary approach should be direct to the study of light nuclei reactions. We have decided to draw our attention  towards this second research line, concentrating on electroweak induced reactions.

The paper is structured in the following way. We will first present the main calculational techniques used to tackle the problem of electroweak reactions with light nuclei in Sec.~\ref{sec:1}.
Then, we will separate the discussion into electromagnetic observables in Sec.~\ref{sec:2} and weak observables in Sec.~\ref{sec:3}. Finally, we will conclude in Sec.~\ref{sec:4} and present an outlook for the future.

\section{Calculational Techniques}
\label{sec:1}

In the investigation of electroweak reactions one fundamental ingredient is the nuclear response
function to the excitation of an external probe. In an {\it inclusive} process it is defined as
\beq
\label{resp}
R_{O}(\omega,q)=\int \!\!\!\!\!\!\!\sum _{f} 
\left|\left\langle \Psi_{f}| O({\bf q})| 
\Psi _{0}\right\rangle\right|
^{2}\delta\left(E_{f}-E_{0}-\omega \right)\,, 
\eeq
where $\omega$ and $q$ are the energy and momentum transferred,  
$| \Psi_{0/f} \rangle$ and 
$E_{0/f}$ denote initial and final state wave functions and energies, and finally ${O}$
represents the excitation operator, which is different according to the external probe.
The $\delta$-function 
ensures energy conservation. 
From Eq.~(\ref{resp}) it is evident that in principle one needs the knowledge of 
all  possible final states excited by the electromagnetic
probe, including of course states in the continuum. Thus, in a straightforward evaluation one 
would have to calculate both 
bound and continuum states. The latter constitute the major
obstacle for a many-body system, since the full  many-body scattering wave functions are not 
yet accessible for A $> $ 3. 
A way to circumvent the problem is to use the Lorentz Integral Transform (LIT) method \cite{EFROS94,REPORT07}.  Within the LIT one considers instead of 
$R_{{O}}(\omega,q)$ an integral transform 
${\cal L}_{{O}}(\sigma,q)$ with a Lorentzian kernel defined for a complex
parameter $\sigma=\sigma_R+i\,\sigma_I$ by
\begin{equation}
{\cal L}_{{O}}(\sigma,q)=\int d\omega
\frac{R_{{O}}(\omega,q)}
{(\omega-\sigma_R)^{2}+\sigma_I^{2}}
= \langle\widetilde{\Psi}_{\sigma,q}^{O}
|\widetilde{\Psi}_{\sigma,q}^{O}\rangle \,.\label{lorentz_transform} 
\end{equation}
The parameter $\sigma_I$ represents the resolution of the transform and is chosen to have 
finite value ($\sigma_I\ne 0$). The basic idea of considering ${\cal L}_{O}$
lies in the fact that it can be evaluated from the norm of a
function $\widetilde{\Psi}_{\sigma,q}^{O}$,
which is the unique solution of the inhomogeneous  equation 
\begin{equation}
({H}-E_{0}-\sigma)|\widetilde{\Psi}_{\sigma,q}^{O}
\rangle={O}(q)|{\Psi_{0}}\rangle\,.\label{liteq}
\end{equation}
Here ${H}$ is the nuclear Hamiltonian.
The existence of the integral in
Eq.~(\ref{lorentz_transform}) implies that  $\widetilde{\Psi}_{\sigma,q}^{O}$ has  
asymptotic boundary conditions similar to a bound
state. Thus, one can apply bound-state techniques for its
solution. 

In all the results presented here, if not indicated differently, in order to
solve Eq.~(\ref{liteq}) we expand $\Psi_{0}$ and $\widetilde{\Psi}_{\sigma,q}^{O}$ in terms
of antisymmetrized hyperspherical harmonics (HH) \cite{paperI}. The expansion is performed up to a  maximal
value of the grand-angular momentum $K_{max}$.
We improve the convergence using
an effective interaction defined within the 
hyperspherical harmonics (EIHH) \cite{EIHH},
where the
starting potential is replaced by an effective one constructed via the Lee-Suzuki method.

The response function $R_{{O}}$ is then obtained by numerically inverting
the integral transform in Eq.~(\ref{lorentz_transform}). For the inversion of the LIT various
methods have been devised~\cite{EfL99,AnL05}.
 In particular the concept and methods of the inversion  have been  recently clarified and
 discussed extensively in Ref.~\cite{ARXIV}, in response to some criticisms. 
Here, we will take care in pointing out
the error bands for the calculated theoretical curves due also to the numerical inversion procedure.

In the next sections we will present recent applications of these calculational techniques to electromagnetic and weak reactions with light nuclei.

\section{Electromagnetic Observables}
\label{sec:2}

\subsection{Photo-absorption Reactions}

In recent years there has been an increased interest in the
photo-disintegration of light nuclei, both in theory and experiment.
From the theoretical side, a lot of effort has been addressed to the investigation of such process
with the LIT. Several light nuclei have been studied with mass number up to $A=7$ \cite{li6he6,li7}. This has stimulated new experimental activity in Lund \cite{lundexp}, for example.
The most systematic calculations with the LIT have been performed on the $^4$He target,
where a variety of realistic potentials have been tested (see e.g.\cite{sonia_solo}, and reference therein).

One important reason for concentrating on $^4$He is that it is the lightest nucleus that
 already possesses features of heavier systems (e.g. large binding 
energy per nucleon). Therefore $^4$He can be considered as an important link 
between the classical few--body systems  and more 
complex nuclei. Furthermore, continuum properties of $^4$He are an 
ideal testing ground for microscopic three-nucleon forces, which are typically fitted 
to bound state properties of few--body systems. 

\begin{figure}[!h]
\centering
\includegraphics[width=0.5\columnwidth,angle=0]{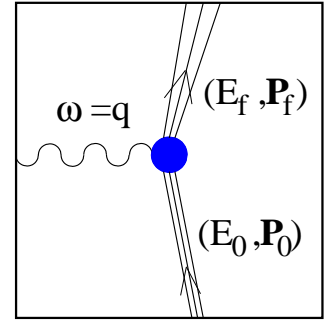}
\caption{Feynman diagram of the photo absorption process.}
\label{diag_photon}       
\end{figure}

The total photo-absorption cross section, represented by the Feynman diagram in Figure \ref{diag_photon}, is given by 
\begin{equation}
\sigma_\gamma(\omega)=4\pi^{2}\alpha\omega R_{{D}_z}(\omega)\,,
\end{equation}
\noindent where $\alpha$ is the fine structure constant and $R_{{D}_z}$ is the response function in  the unretarded dipole
approximation 
\begin{equation}
{\label{1}
R_{{D}_z}(\omega )=\int d\Psi _{f}\left| \left\langle \Psi _{f}\right| {D}_z \left| 
\Psi _{0}\right\rangle \right| ^{2}\delta (E_{f}-E_{0}-\omega)} \,,
\end{equation}
with
\begin{equation}
{D}_{z}=\sum_{k=1}^{A}\frac{\tau^{3}_{k}z_{k}}{2}\,,
\end{equation}
being the unretarded dipole operator. Here,  
$\tau^{3}_{k}$  and $z_{k}$ 
represent the third component of the isospin operator and of the spatial
coordinate of the $k$-th particle in the center of mass frame.
This is a typical example of an electromagnetic induced reaction that can be 
studied with the LIT.

Photo-nuclear processes can be of great 
importance, since exchange currents, which are connected to the nuclear 
interaction by current conservation, play a decisive role in the cross section. 
Within the unretarded dipole approximation,
 the dominant part of the  exchange
current is taken into account implicitly via the
Siegert theorem, particularly for 
photon energies $\omega$ below 50 MeV. This has been shown for the deuteron \cite{SaA}
and is expected to be valid for $^4$He as well.

\begin{figure}[!htb]
\centering
\includegraphics[width=1.0\columnwidth,angle=0]{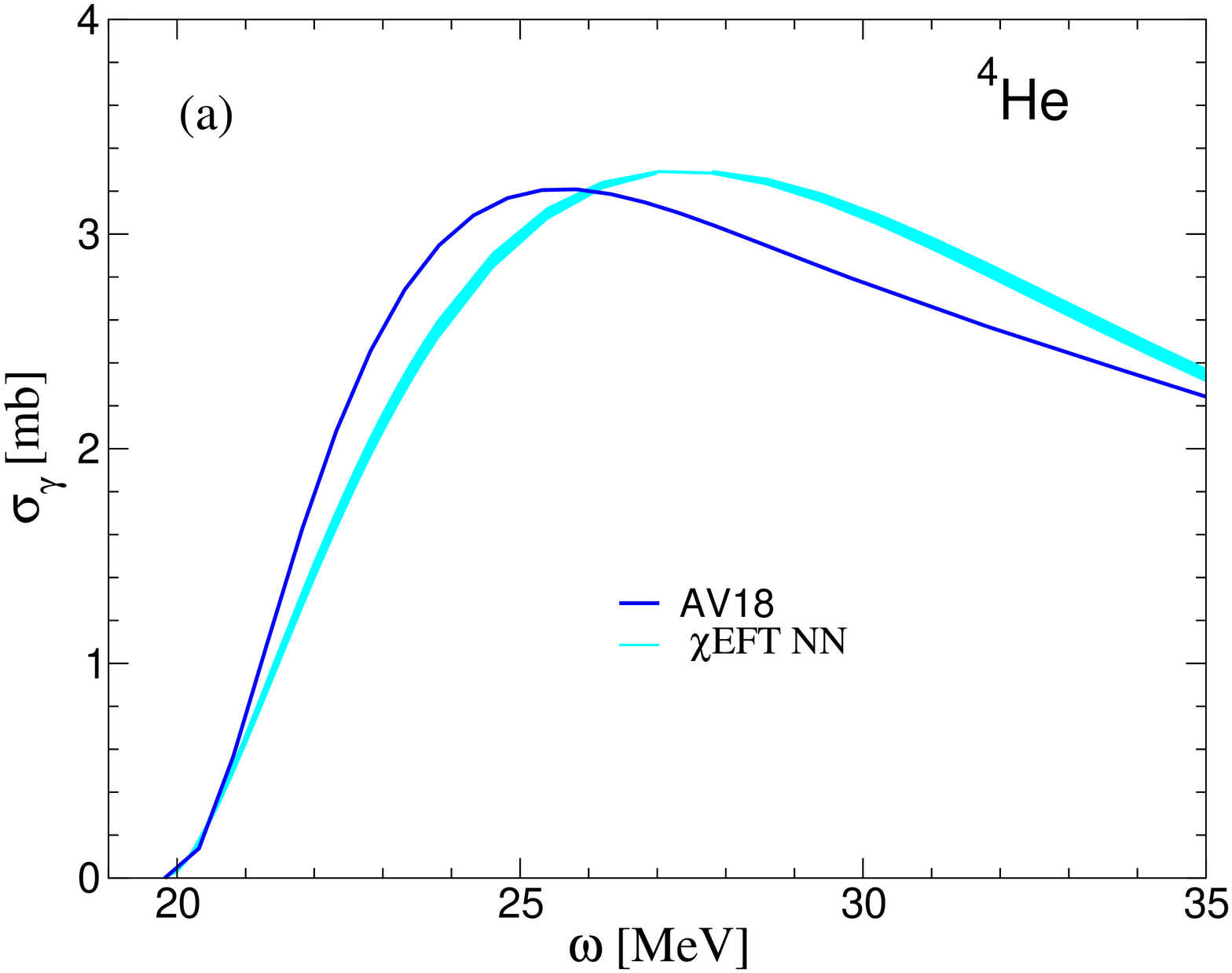}
~\\
~\\
\includegraphics[width=1.0\columnwidth,angle=0]{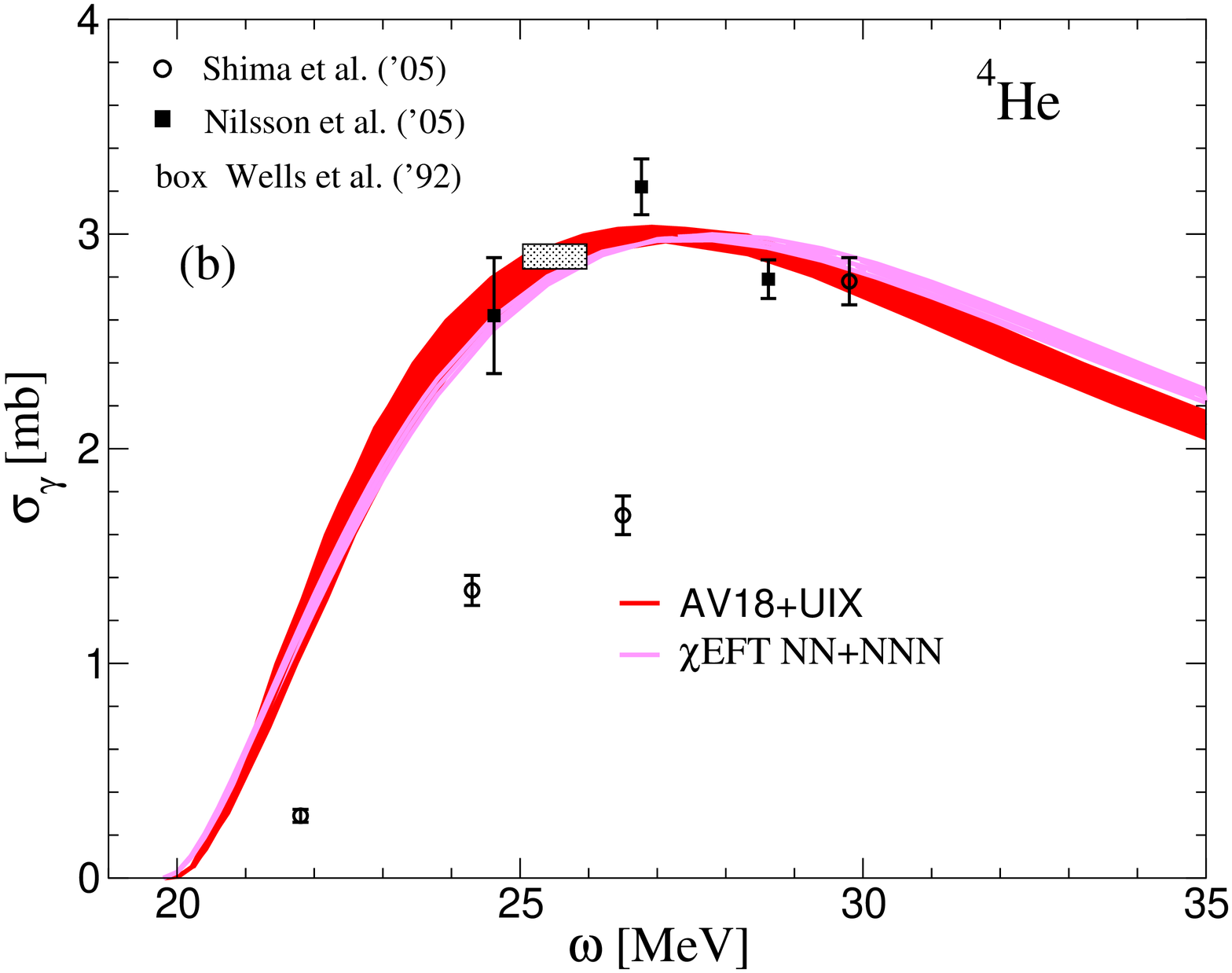}
\caption{(a) Total $^4$He photo-absorption cross section $\sigma_\gamma$ for AV18 
(dark blue) and the  $\chi$EFT N$^3$LO (light blue) potentials; 
(b) $\sigma_\gamma$ including the 3NFs UIX (red) and the  $\chi$EFT N$^2$LO (pink) in comparison with  with some experimental data:  box from \cite{Wells}, squares from
\cite{Lund}, and circles from \cite{Shima}.}
\label{fig:1}       
\end{figure}

The first realistic calculation of the $^4$He photo-absorption was performed
with the LIT and EIHH method \cite{Gazit06} using the Argonne $V_{18}$ (AV18) \cite{AV18} 
potential augmented by an Urbana IX (UIX) \cite{UIX} three-nucleon force. Later on, a second
realistic calculation was performed with the LIT and the no-core shell model (NCSM) \cite{SofiaPhoton}  using chiral two- and three-body forces \cite{EM}.
It is very interesting to compare these two calculations in the following way.
First one can look at the predictions  from  two-body forces only. Figure \ref{fig:1}(a) 
shows in fact the result for $\sigma_\gamma(\omega)$ with the AV18 and the $\chi$EFT force at N$^3$LO \cite{EM}.
One observes that, even though both forces reproduce very well the nucleon-nucleon (NN) experimental phase shifts (with a $\chi^2$ per datum of about 1), the prediction of this many-body observable
is different, due to the different off-shell properties of the two-potentials, which are not fixed by the NN phase-shifts.
 Figure \ref{fig:1}(b) 
shows, instead,  $\sigma_\gamma(\omega)$ calculated from Hamiltonians that include a three nucleon force:
in one case the UIX and in the other the  $\chi$EFT force at N$^2$LO. We observe that the two predictions are very similar, even though the 3NF have been constrained differently (the UIX on the triton binding energy and on saturation properties of nuclear matter and the chiral N$^2$LO  on $s$- and $p$-shell nuclei \cite{Navratil07}). 

One should note that the theoretical error bands shown in Fig.~\ref{fig:1} take into 
account the numerical error of the LIT inversion. Furthermore, the AV18+UIX curve has
a larger width due to the fact that an extrapolation was done to an infinite model space for the LIT itself \cite{Gazit06}.   

It is particularly interesting to observe  the relevance of 3NFs. 
Due to  3NFs one observes a reduction of the peak height by less than 10\% both 
for  the more phenomenological potentials as well as in the chiral approach. One should note, though, that larger 3NF effects were found with the UIX in \cite{Gazit06} at higher energies than those shown in Fig.~\ref{fig:1}.

In Fig.~\ref{fig:1}(b) we also compare to some selected experimental data (for more data see \cite{Gazit06} and \cite{SofiaPhoton}): the box is from \cite{Wells}, where the peak cross section is 
determined from Compton scattering via dispersion relations;
 squares from \cite{Lund}  and dots from \cite{Shima} show  measurements of the cross section in a broader energy range.
It is evident that the various experimental cross 
sections are quite different exhibiting maximal deviations of about a factor of 
two. While theory is relatively accurate, considering the sum of the two theoretical bands in Fig.~\ref{fig:1}(b) as an estimate of the total theoretical error,  the experimental 
situation is unsatisfactory and does not allow one to draw any definitive conclusions on the importance of 3NFs to describe this observable.

\subsection{Electron Scattering Reactions}

The information about the nuclear dynamics that one can get from electron scattering
processes is complementary to the one obtained with real photons and much richer. While the energy of the photon fixes the momentum transferred to the
target, an electron can transfer energy $\omega$ and momentum $q$ independently.
As  $q=|{\bf q}|$ increases one can probe  properties of the correlations among nucleons from the long- to short-range.
\begin{figure}[!h]
\centering
\includegraphics[width=0.6\columnwidth,angle=0]{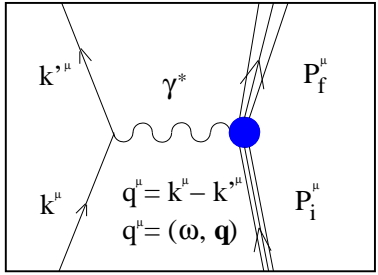}
\caption{Feynman diagram of the electron scattering process at the one-photon exchange level.}
\label{diag_el}       
\end{figure}

Various {\it inclusive} experiments have already been performed in the past on light nuclei. Due to the low atomic number it is possible to study the longitudinal and the transverse responses without the ambiguities created by Coulomb distortions affecting heavier systems.

We report on recent studies of the the {\it inclusive} inelastic electron scattering reactions on light nuclei with $A=$4 and 3 on a broad energy range obtained using 
the LIT method. This is, in fact, the only way to allow investigations beyond the three- and four-body (for $A=4$) break up thresholds.

In the one-photon-exchange approximation (see Fig.\ref{diag_el}), the inclusive cross section
for electron scattering off a nucleus is given in terms of two
response functions,  
\begin{equation}
\frac{d^2 \sigma}{d\Omega d{\omega}}=\sigma_M\left[\frac{Q^4}{q^4}
{R_L(\omega,q)}+\left(\frac{Q^2}{2 q^2}+\tan^2{\frac{\theta}{2}}\right)
{R_T(\omega, q )}\right]\,,
\end{equation}
where $\sigma_M$ denotes the Mott cross section, $Q^2=-q_{\mu}^2={
q}^2-\omega^2$, and
$\theta$ the electron scattering angle. The longitudinal 
and transverse response functions,
$R_L(\omega,{q})$ and $R_T(\omega,{ q})$, are determined
by the transition matrix elements of the Fourier transforms  of the 
charge and the transverse current density operators, respectively.
Experimentally $R_L$ and $R_T$ can be disentangled by using 
a Rosenbluth separation method.

We will start presenting a study of the longitudinal response function for $^4$He and $^3$H and
then continue with the transverse response function for $^4$He and the $A=3$ nuclei. 

\subsubsection{The Longitudinal Response Function}

The longitudinal response function is given by
\beq
\label{frisp}
R_L(\omega,q)=\int \!\!\!\!\!\!\!\sum _{f} 
\left|\left\langle \Psi_{f}| \rho({\bf q})| 
\Psi _{0}\right\rangle \right|
^{2}\delta\left(E_{f}+\frac{q^2}{2M}-E_{0}-\omega \right)\,, 
\eeq
where  
$M$ is the target mass. The charge density operator $\rho$ is defined as
\beq
{\rho}({\bf q})= 
\sum_k  e_k \exp{[i {\bf q} \cdot {\bf r}_k]} \,, \label{rho}
\eeq
where $e_k$ is the nucleon charge in unit of $e$\footnote{$e$ is here factorized in $\sigma_M$.}
 and ${\bf r}_k$ is the coordinate of the $k$-th particle in
the center of mass frame.
In order to take the finite size of the nucleon into account, form factors are introduced, such
that the nucleon charge takes the following form
\begin{equation}
\label{ek}
e_k=\frac{1+\tau^3_k}{2}G^p_E(Q^2)+ \frac{1-\tau^3_k}{2}G^n_E(Q^2) \,,
\end{equation}
where $G^{p/n}_E$ are the proton and neutron electric form factors. We use the usual dipole fit 
for them \cite{GaK71}.

\begin{figure}[!htb]
\centering
\includegraphics[width=1.0\columnwidth,angle=0]{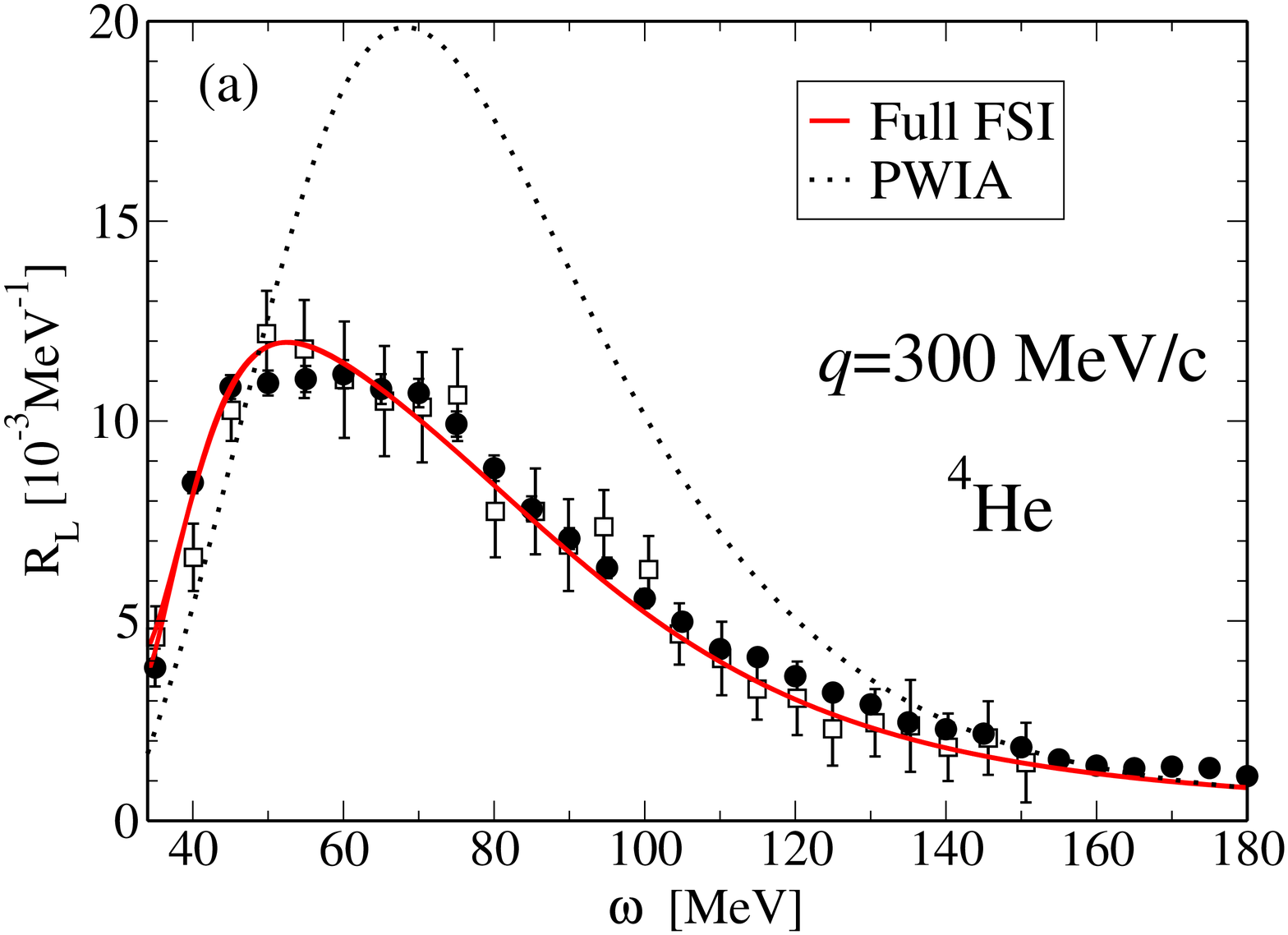}
~\\
\includegraphics[width=1.0\columnwidth,angle=0]{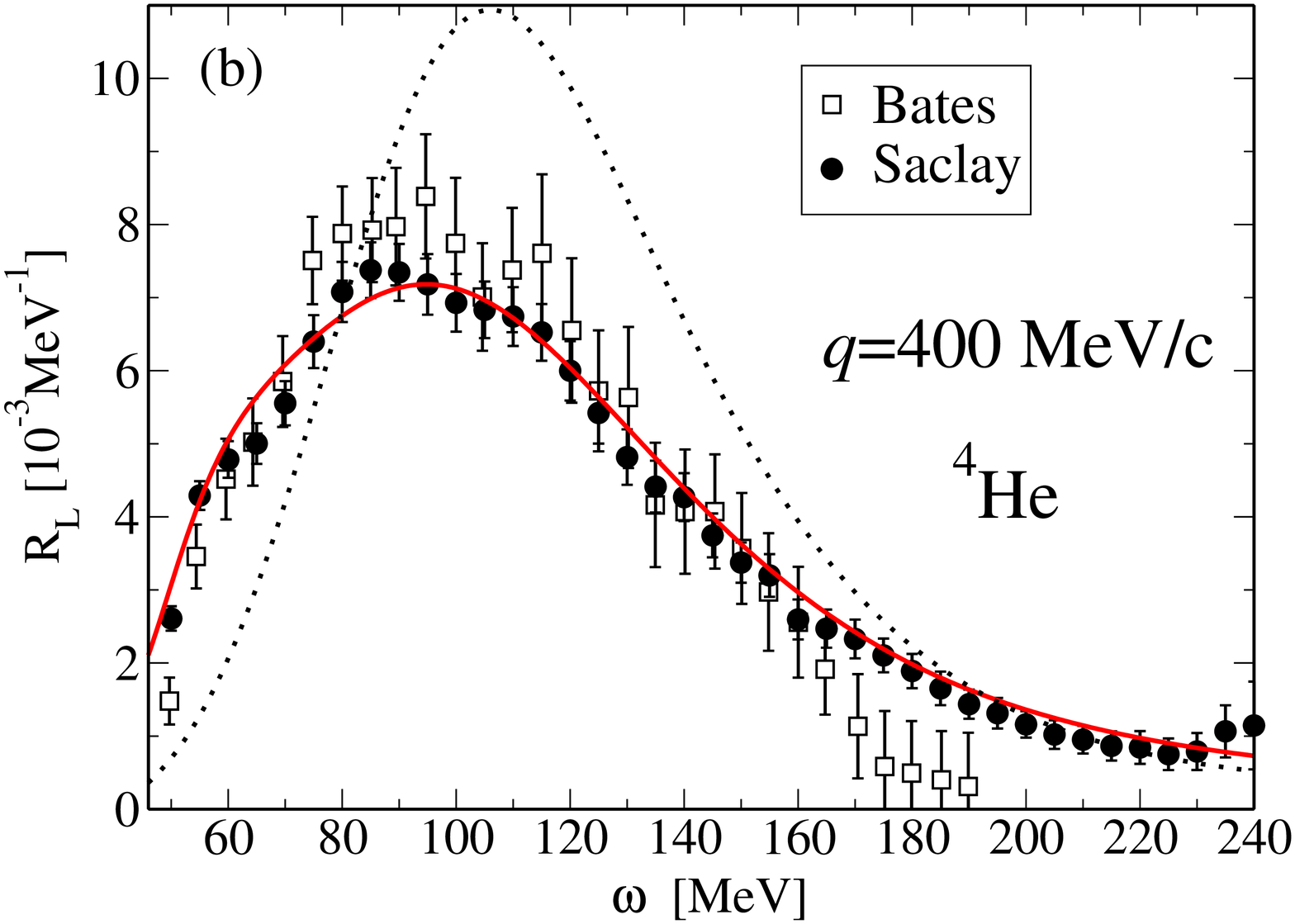}
~\\
\includegraphics[width=1.0\columnwidth,angle=0]{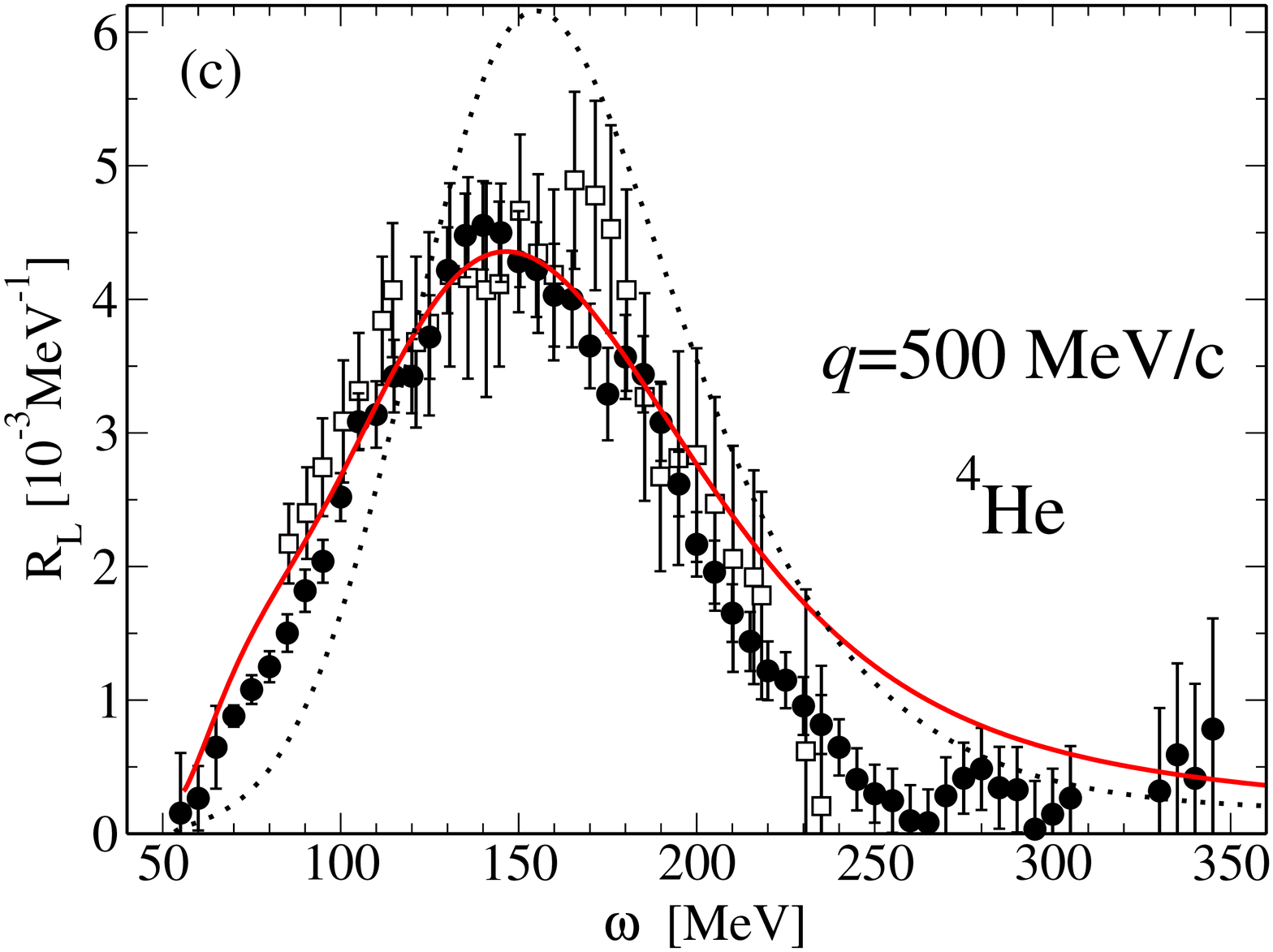}
\caption{Longitudinal response function for $^4$He at various $q$: PWIA (dotted) versus the calculation including the full FSI (solid) from the AV18+UIX potential. Data shown are from Bates~\cite{Bates} (squares), Saclay~\cite{Saclay} (circles).}
\label{fig:2}       
\end{figure}

Since the longitudinal response $R_L$ is much less sensitive to meson exchange
 effects  than the transverse response  
$R_T$,  the use of a simple one-body 
density operator allows one to concentrate on the nuclear dynamics generated by the potential.
In fact, for low $q$ two-body operators in $R_L$ are only of fourth order
in effective field theory counting (N$^3$LO) \cite{PARK03}, and their
contribution is negligible up to $q \approx$ 300 MeV/c.
In recent works \cite{PRL09,PRC09} we have studied the longitudinal response function $R_L(\omega,q)$ for $^4$He at constant momentum transfers $q\le$ 500 MeV/c. 
We consider the non-relativistic charge  density operator of Eq.~(\ref{rho}), and  
decomposed it into isoscalar and isovector Coulomb multipoles \cite{PRC09}. 
Each multipole is then used as source in the right-hand-side of Eq.~(\ref{liteq})
which is solved with the EIHH method. After inversion of each multipoles and a subsequent sum
of all of them, one obtains results that can be directly compared with experiment.
The accuracy of the convergence for the LIT is found to be at a percentage level for all the results shown in the next paragraphs, including the numerical error of the inversion.

It is  first interesting  to investigate  the role of the final state interaction (FSI) in the electron scattering reaction.
In \cite{Euclidean} it was found that the FSI is very important for $R_L$ in the quasi-elastic peak of $^4$He.
To investigate that we compare our LIT results using the AV18+UIX potential both in the initial and final state with a plane wave impulse approximation (PWIA) calculation. 
The PWIA result is obtained under the hypothesis  of  one outgoing free proton with mass $m$ and a spectator (A-1)-system with mass $M_s$:
\begin{equation} 
  R_L^{PWIA}(\omega,q)=\! \int d{\boldsymbol p}  \,n({\boldsymbol p})\, 
  \delta \left( \omega-\frac{( {\boldsymbol p} + {\boldsymbol q} )^2}{2
  m}-\frac{\bs{p}^2}{2 M_{s}} - \epsilon \right). 
\end{equation}
Here $n({\boldsymbol p})$ represents the proton momentum distribution and
$\epsilon$ the proton separation energy.  To obtain a PWIA curve we have used the $n({\bf p})$ of ~\cite{Wiringa}. 
In Fig.~\ref{fig:2}, the results of $R_L(\omega,q)$ with the AV18+UIX potential at various $q$ are shown and
compared to data. In all cases one finds that the FSI effects are very large
and essential for reaching agreement with experiment, confirming what was previously found. One has to note, that
the PWIA fails particularly 
in the q.e. peak and at low $\omega$, while it is better on the tail.

\begin{figure*}[!htb]
\centering
\includegraphics[width=1.0\columnwidth,angle=0]{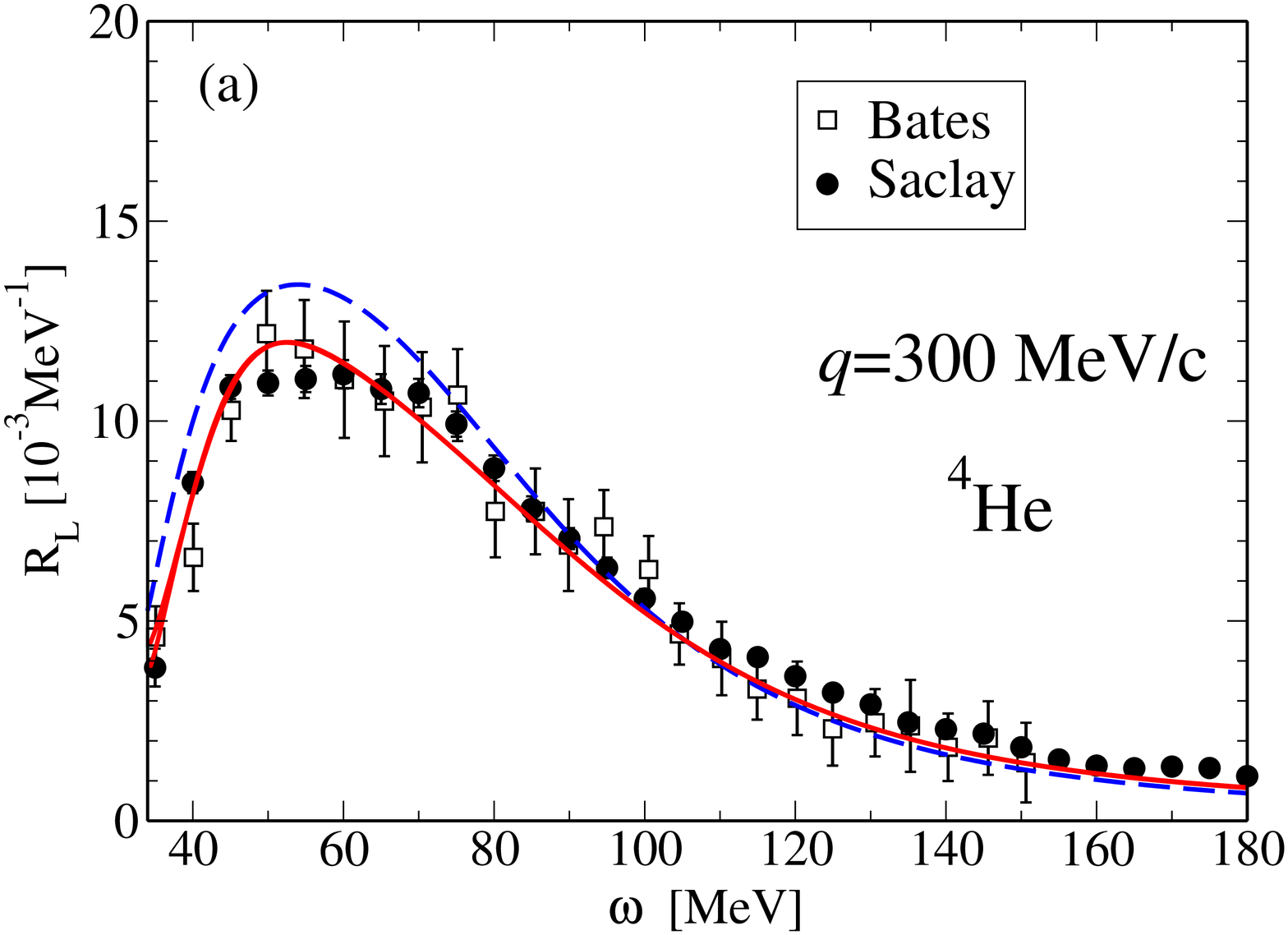}
\includegraphics[width=1.0\columnwidth,angle=0]{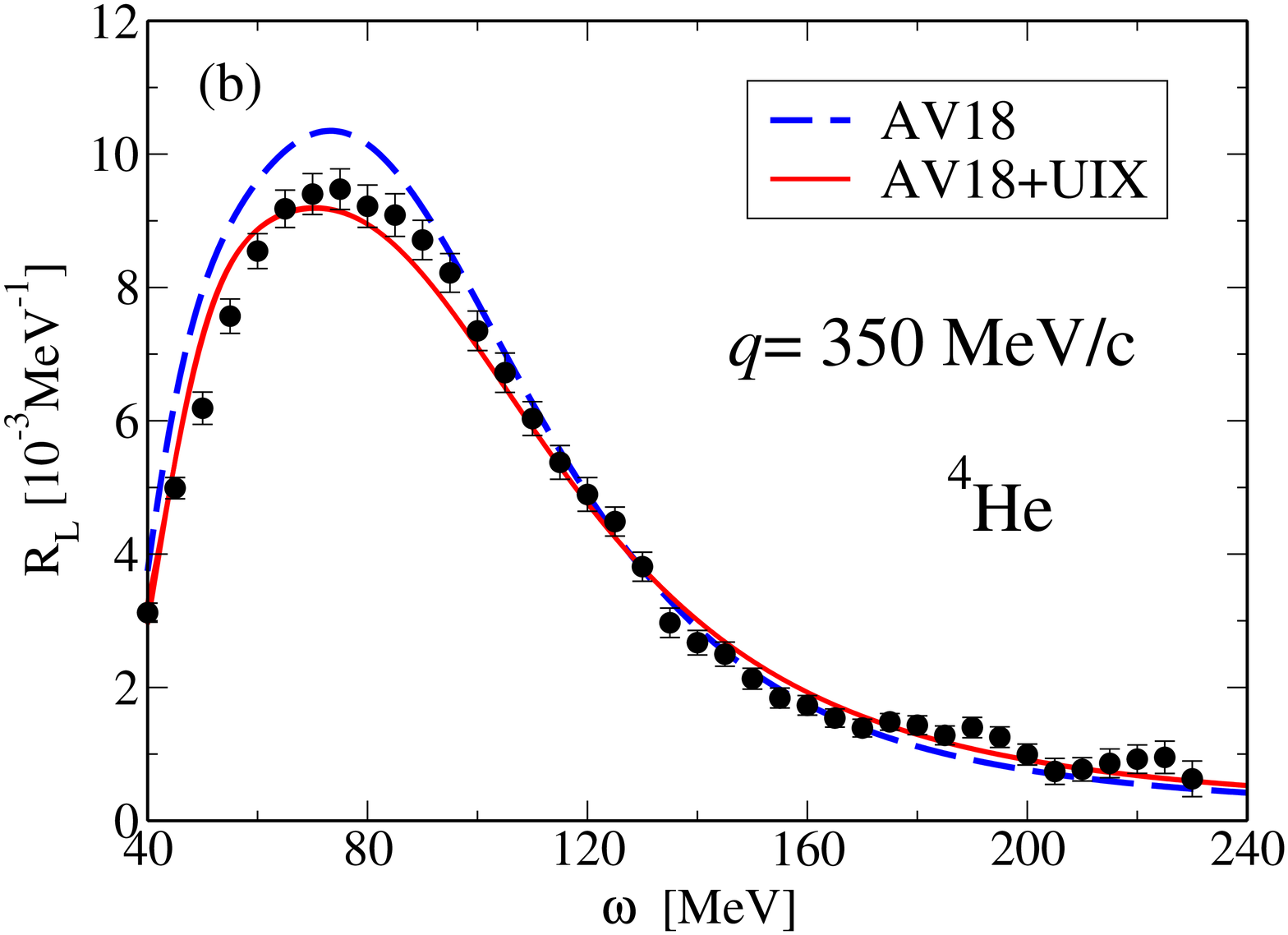}
~\\
\includegraphics[width=1.0\columnwidth,angle=0]{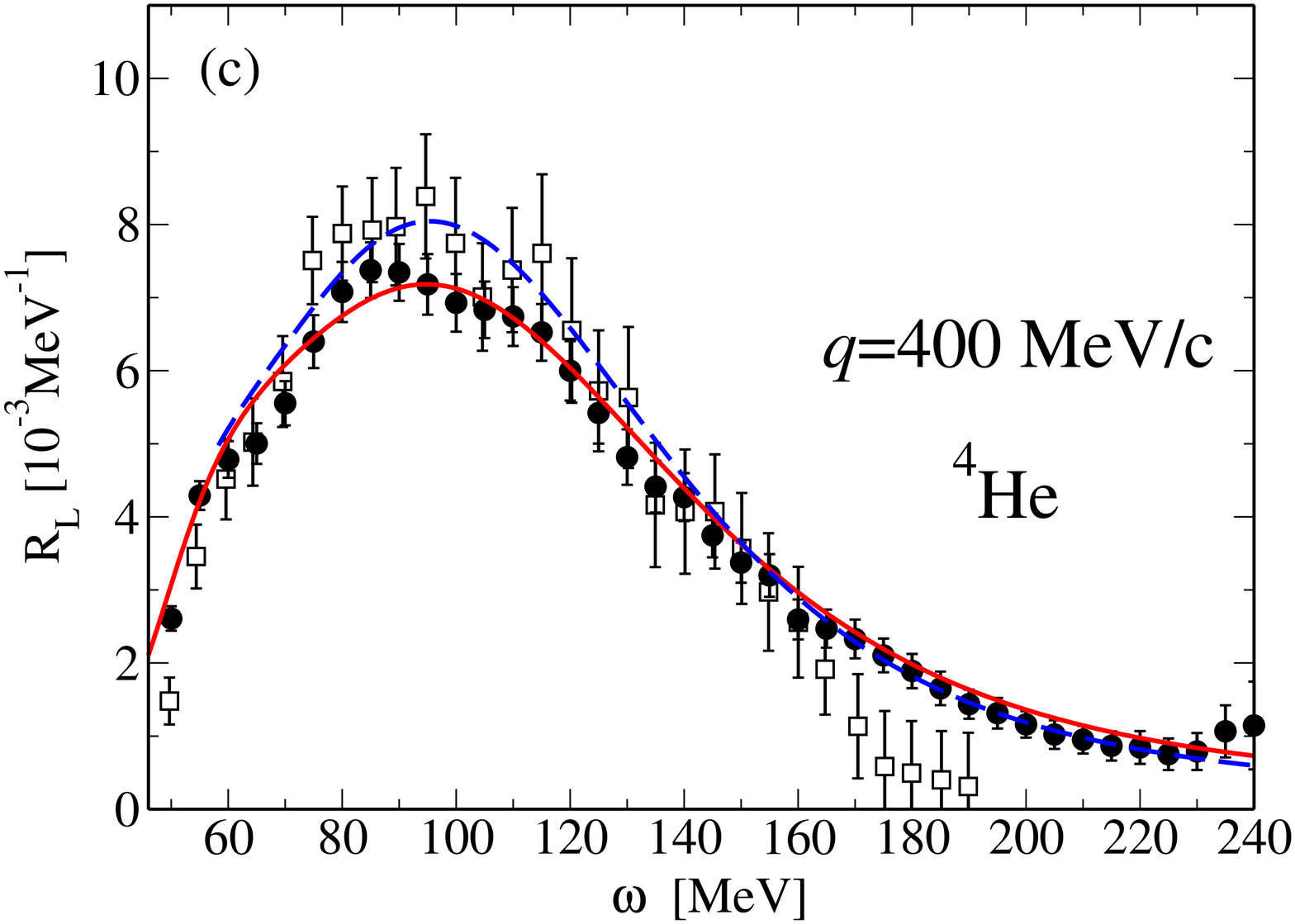}
\includegraphics[width=1.0\columnwidth,angle=0]{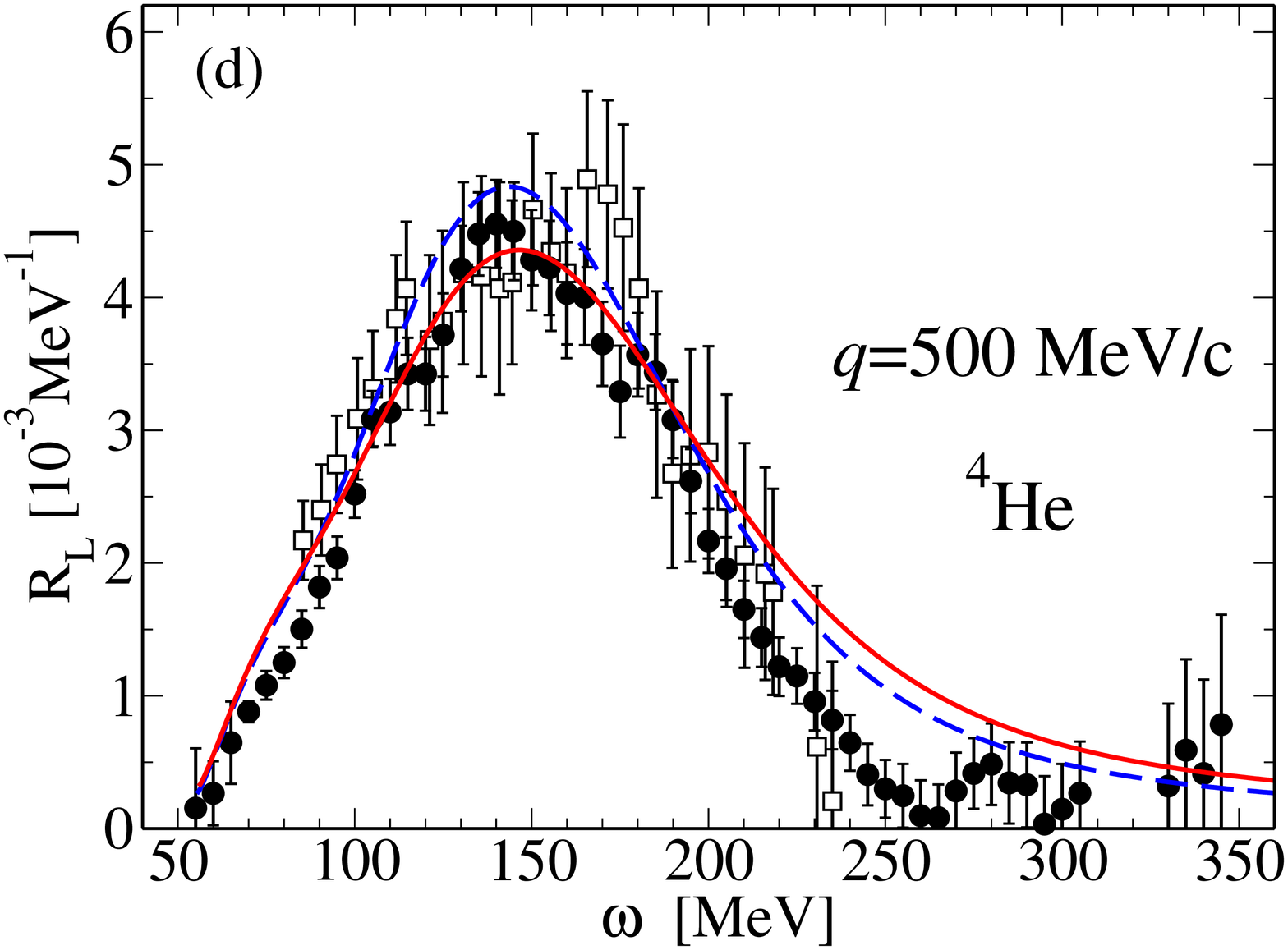}
\caption{Longitudinal response function at various $q$: Calculation with the AV18 (dashed) and with the AV18+UIX (solid) potential in comparison with experimental data from Bates~\cite{Bates} (squares), Saclay~\cite{Saclay} (circles).}
\label{fig:3}       
\end{figure*}

The second fact one can investigate is the role of three-nucleon forces on this observable.
To this aim, we can for example compare calculations with a two-body force only, like the AV18 potential, with the one with AV18+UIX.
In Fig.~\ref{fig:3}, we show  this  comparison for some values of $q$ together with existing experimental data. One sees that the 3NF results are closer to experiment, this 
is particularly evident at $q=300 $ and $350$  MeV/c. The 3NF effect is generally not very large, but always leads to a quenching of the strength. For some $q$ values the 3NF quenching is  
comparable to the size of the experimental error bars, particularly for the data from Ref.~\cite{Bates}.
The largest discrepancies with data are found at $q=500$ MeV/c.
While the height of the peak is well reproduced by the result with 3NF, the width
of the experimental peak seems to be somewhat narrower than the theoretical one. On the other hand
 one has to be aware that relativistic effects, which we are not including,
are not completely negligible at such momentum transfers (we will comment on that later).

\begin{figure}[!htb]
\centering
\includegraphics[width=1.0\columnwidth,angle=0]{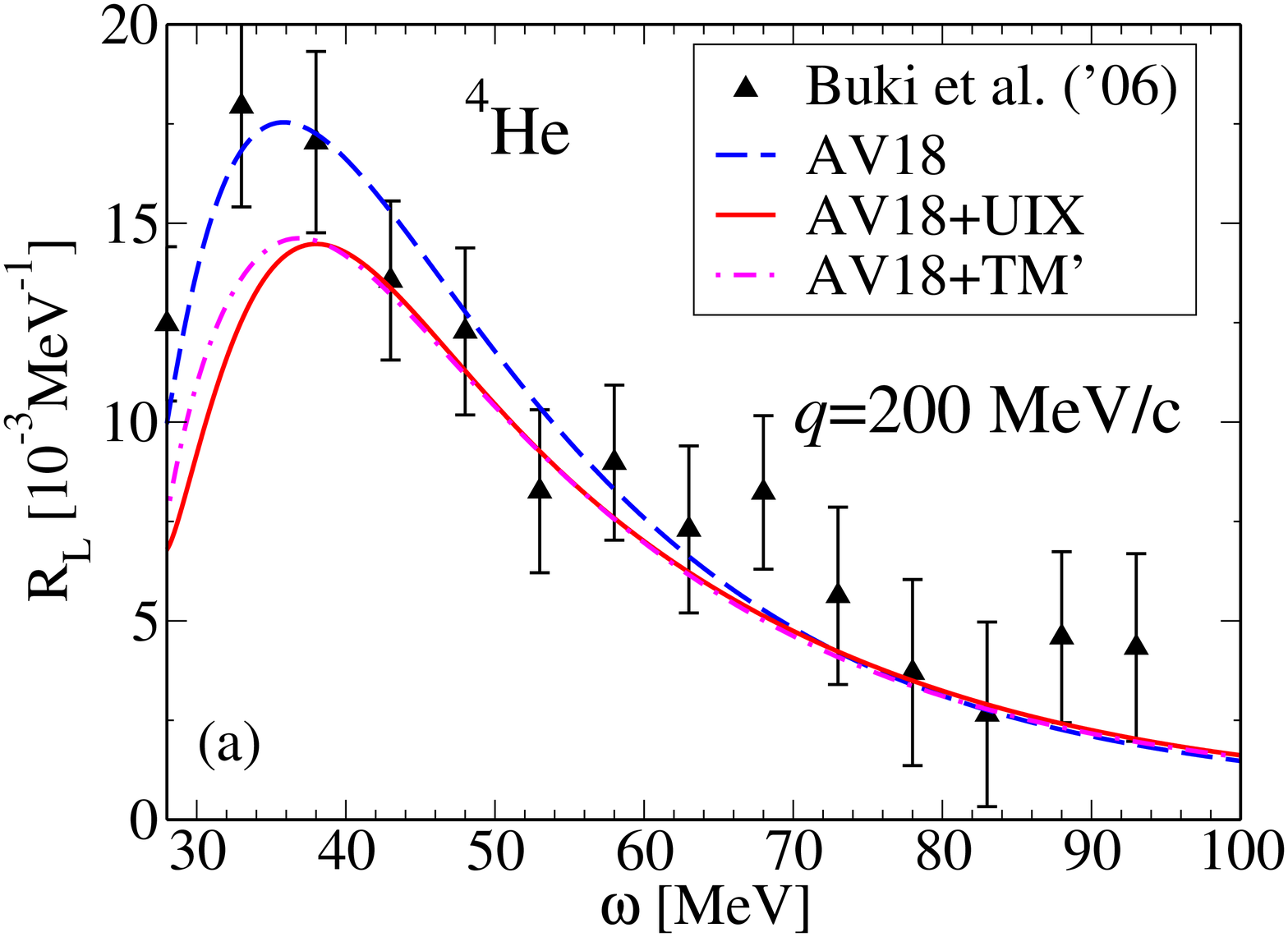}\\
~\\
\includegraphics[width=1.0\columnwidth,angle=0]{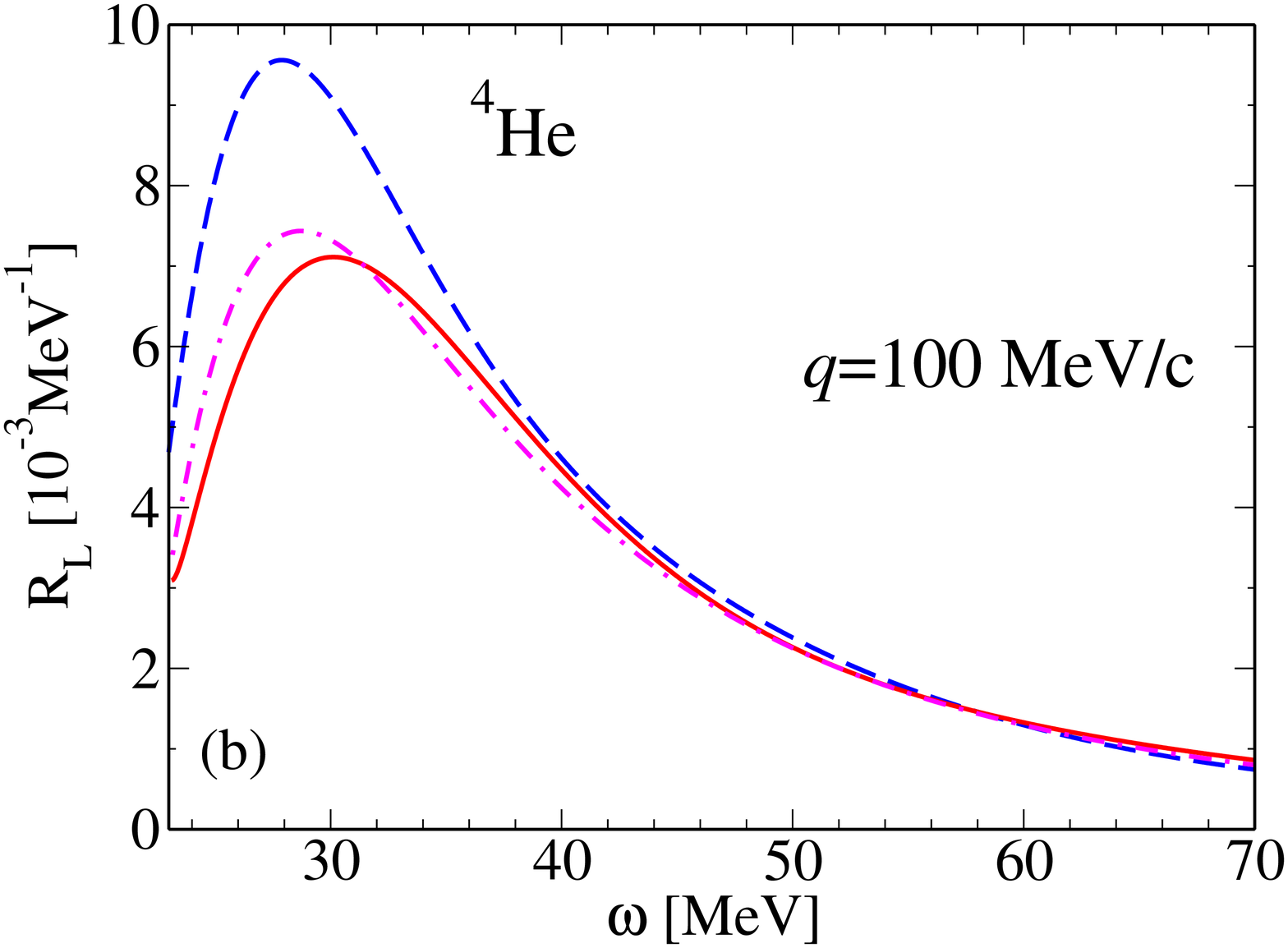}\\
~\\
\includegraphics[width=1.0\columnwidth,angle=0]{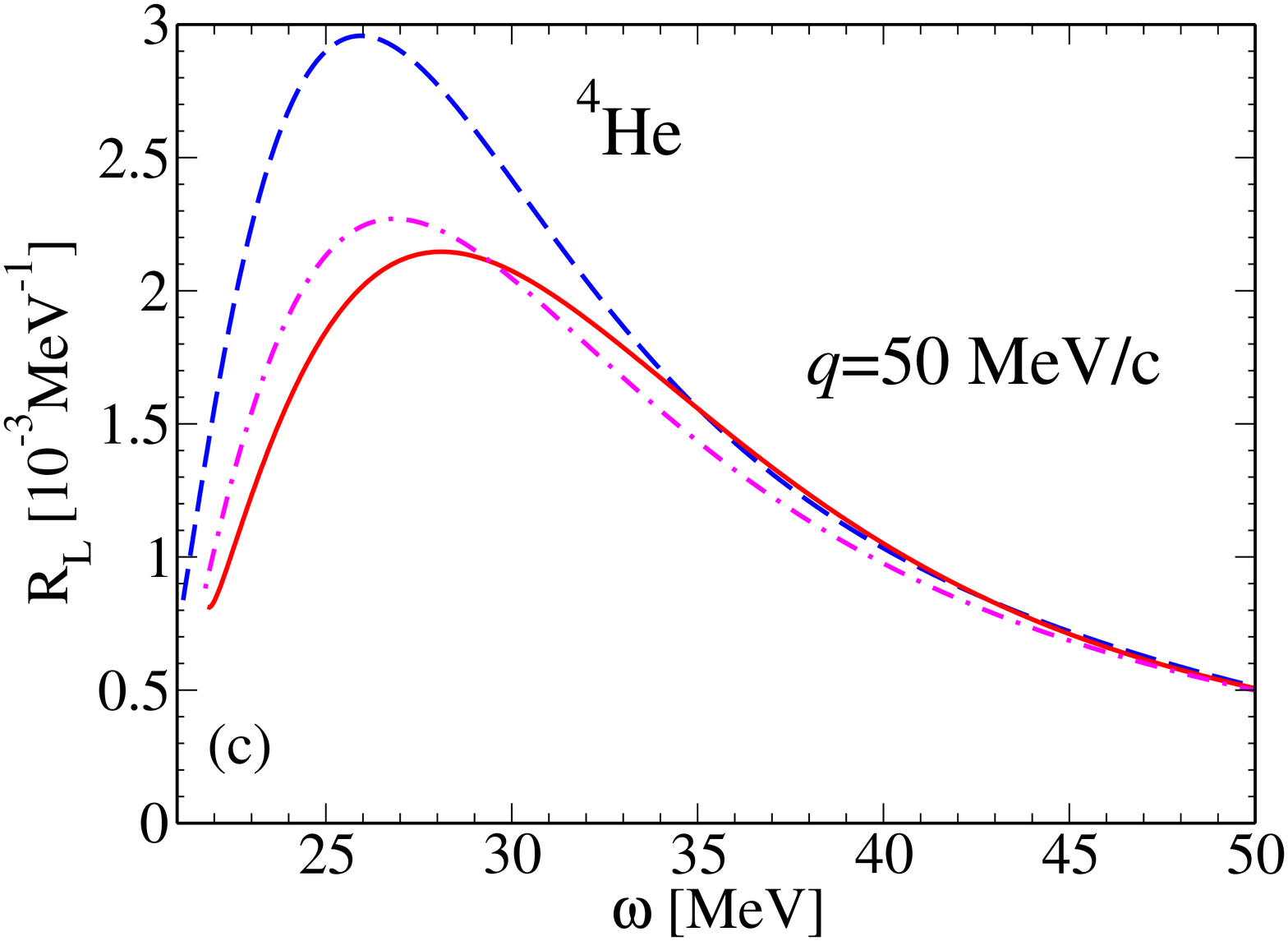}\\
\caption{Longitudinal response function for low-momentum transfer 
 with  the AV18 two-nucleon force only (dashed),  and with the addition of the UIX (solid) or the TM' (dashed-dotted) three-nucleon force.}
\label{fig:4}       
\end{figure}

It is worth looking at the 3NF effect when even lower momentum transfers are considered,
even though for some kinematics there are no experimental data to compare to.
In Fig.~\ref{fig:4}, we show $R_L$ at three  momentum transfer $50 \le q \le 200$ MeV/c. 
We find a large quenching effect due to 3NFs, which 
is strongest at lower  $q$. We investigate it  with two different 3NF models: the UIX and the
 Tucson Melbourne (TM')~\cite{TMprime}, when starting from the same two-nucleon AV18 potential. 
While the UIX force contains a two-pion exchange and a short range phenomenological term,
the TM' force includes two-pion exchange terms where the coupling constants are taken from pion-nucleon scattering 
data consistent with chiral symmetry. 
The cutoff of the TM' force has been calibrated on triton binding energy, when used in conjunction with the AV18 NN force. Thus, the TM' is not adjusted also to nuclear matter properties as the UIX. 
With a cutoff $\Lambda=4.77m_\pi$, where $m_\pi$ is the pion mass, we obtain the following binding energies
8.47 MeV ($^3$H) and 28.46 MeV ($^4$He) for the AV18+TM' potential. We  would like to emphasize that the $^4$He binding energy 
is practically the same as for the AV18+UIX case (as already found in \cite{Nogga02}).

Figure~\ref{fig:4} shows that the increase of 3NF effects 
with decreasing $q$ is found for both the UIX and TM' force model.
The quenching of the strength is rather strong,  reaching  
$40\%$ for $q=50$ MeV/c.
 Moreover
it becomes evident that  the difference between the results obtained with the two  3NF models 
increases  with decreasing $q$. One actually finds that
the shift of the peak to higher energies in the case of UIX  generates for $R_L$ a difference up to about 10\%
on the low
energy sides of the peaks. This is very interesting, since it represents the  case of an 
electromagnetic observable whose prediction depends on the choice of the 3NF.
In  light of these results it would be very interesting to repeat the calculation with  EFT 
two-and three-body potentials \cite{EM,EGM}.
 While to our knowledge there are not available measurements at the low momentum kinematics like $q=50$ and $100$ MeV/c,
recently some data have been taken at $q \simeq 200$ MeV/c \cite{Buki06}.
They are shown in Fig.\ref{fig:4}. While one finds a satisfactory agreement between the AV18+UIX
result and data beyond the peak, one observes a non negligible discrepancy at the peak itself.
Moreover the error bars are quite large.  
It would be highly desirable to have precise measurements of $R_L$ at low $q$.
 This could serve either to fix the 
low-energy constants (LEC) of the effective field theory three-body force or to possibly discriminate between different
nuclear force models. Here, it is worth mentioning that our calculations have stimulated new
experimental activity in Mainz \cite{Distler}.
More experiments on the same observable  also planned at 
Jefferson Laboratory (E05.110 at Hall A) at large momentum transfer. 
 
\begin{figure}[!htb]
\centering
\includegraphics[width=1.0\columnwidth,angle=0]{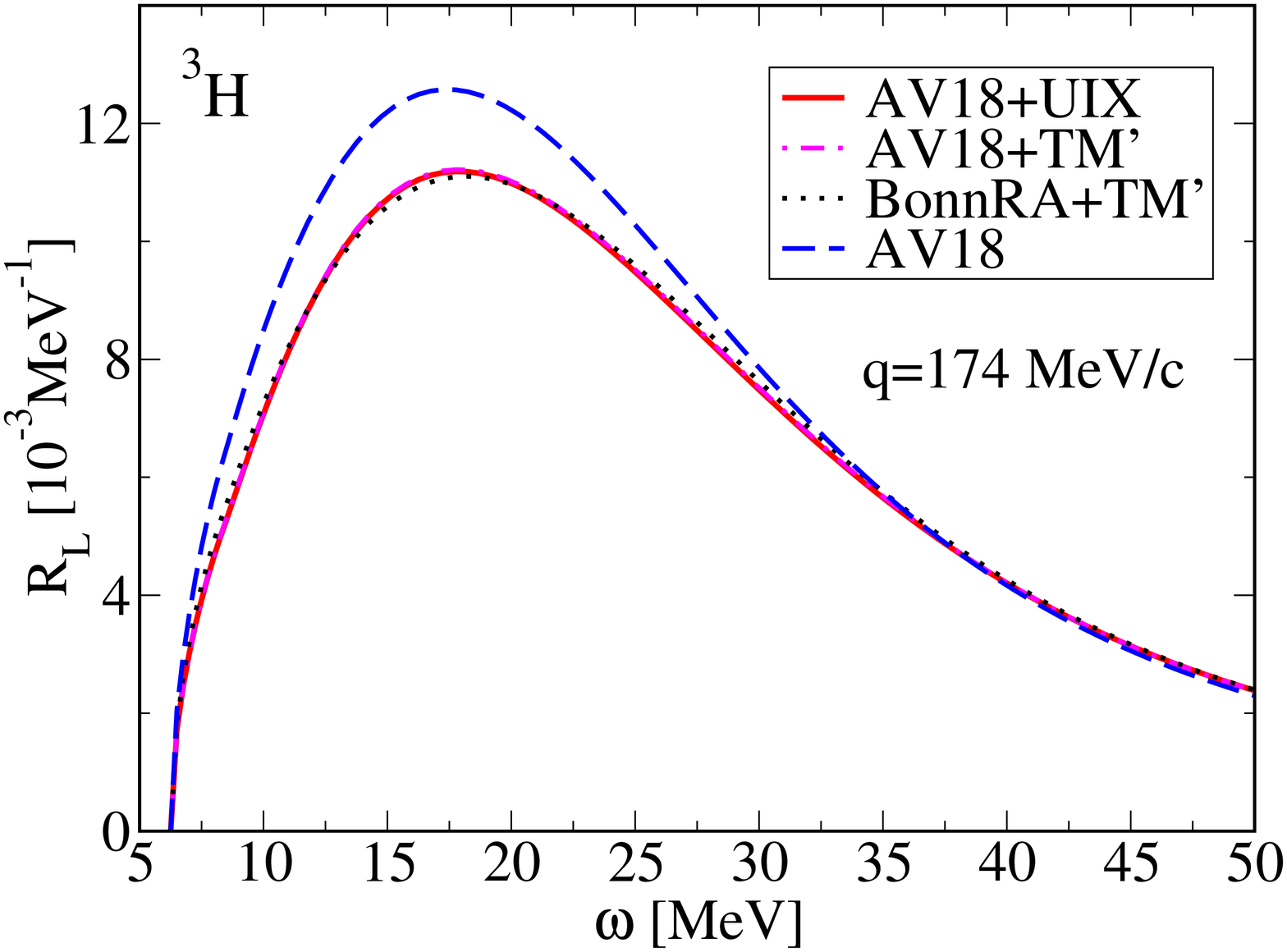}
\caption{Longitudinal response function of $^3$H for $q=174$ MeV/c 
 with  the AV18 two-nucleon force only (dashed),  and with the addition of the UIX (solid) or the TM' (dashed-dotted) and with the Bonn two-body potential with the TM' 3NF (dotted). Calculations are taken from \cite{RLA3}.}
\label{fig_RLA3}       
\end{figure}

It is interesting to compare the $^4$He longitudinal response at low momentum transfer with the corresponding 
calculations done for $A=3$ nuclei published in \cite{RLA3}. Also for $A=3$ nuclei a reduction of the peak due to the 3NF was observed and it was found to be more pronounced at low momentum transfer. Though the percentage effect is different for $A=4$ and $A=3$ nuclei.
In case of the $\alpha$-particle the 3NF force is quenching the quasi-elastic peak of about $20\%$ for $q=200$ MeV/c. For $^3$H, instead, the quenching effect of 3NF was found to be of about $10\%$ only at $q=174$ MeV/c at energies between 10 and 20 MeV. This can be seen in Fig.\ref{fig_RLA3}, where theoretical results with the same potentials used for $^4$He are shown. In addition, a calculation with the configuration space BonnA \cite{BonnRA} potential augmented by the TM' force was done. Comparing the three cases in Fig.\ref{fig_RLA3} where a 3NF was included, one finds that
the result is very stable.

 Summarizing, in case of $^4$He two features are evident with respect to $^3$H: (i) there is a larger effect of 3NFs and (ii) the two different three-body force models lead to two different curves.
This shows that $^4$He is a more interesting target to study 3NFs, since their effect is amplified. This is expected since it is a denser systems and 
the ratio between the number of triplets and of pairs goes like (A-2)/3, therefore doubles from 
$^3$H to $^4$He.

Having the full response function at our disposal, we can also calculate some integral properties to compare with other calculations found in literature.
The classical sum rule of $R_L(\omega,q)$, that has been much discussed in the literature, is the Coulomb sum rule (CSR) \cite{REP91}. It consists in  the integral of the inelastic 
longitudinal response function
\begin{eqnarray}
\nonumber
{\rm CSR}(q)&=&\int_{\omega_{th}}^{\infty}\!\!\!\!\!\! d\omega
{\frac{R_L(\omega,q)}{|G_E^p(Q^2)|^2}}\\
& =& Z  + Z (Z-1) f_{pp} (q) -Z^2 |F(q)|^2
\!,\label{CSR} 
\end{eqnarray}
that can be connected to 
 the number of protons $Z$ and
 to  $f_{pp} (q)$, which  is the Fourier 
transform of $\rho_{pp}(s)$,  i.e the probability to find two 
protons at a distance $s$.
Here $F(q)$ is the nuclear elastic form factor.  

\begin{table}
\caption{CSR of Eq.~(\ref{CSR}) for different momentum transfers obtained with the AV18 and the AV18+UIX potential.}
\label{tab:1}       
\begin{tabular}{ccc}
\hline\noalign{\smallskip}
q [MeV/c] &CSR with AV18 & CSR with AV18+UIX  \\ 
\noalign{\smallskip}\hline\noalign{\smallskip}
50    &  0.064  &   0.053    \\ 
100   &  0.239  &   0.211    \\ 
150   &  0.516  &   0.460    \\ 
200   &  0.839  &   0.771    \\ 
250   &  1.163  &   1.072    \\ 
300   &  1.437  &   1.350    \\ 
350   &  1.634  &   1.562    \\ 
400   &  1.771  &   1.717    \\ 
450   &  1.842  &   1.816    \\ 
500   &  1.884  &   1.871    \\ 
\noalign{\smallskip}\hline
\end{tabular}
\end{table}
\begin{figure}[!htb]
\centering
\includegraphics[width=1.0\columnwidth,angle=0]{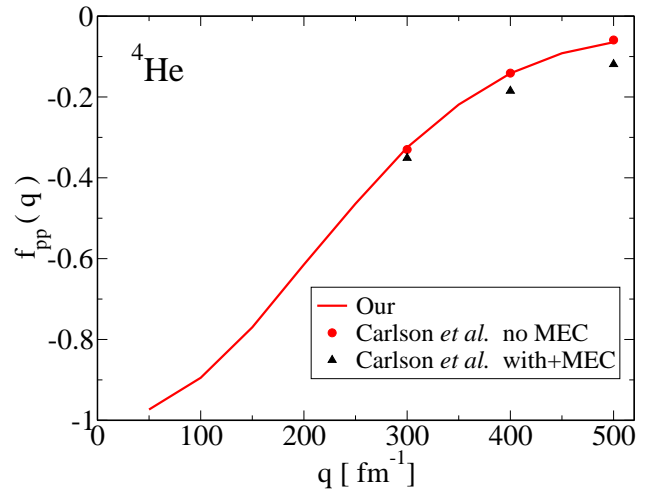}
\caption{Our calculation of $f_{pp}(q)$ for $^4$He 
with  AV18+UIX (solid) potentials, in comparison with
the results from~\cite{SICK} with the one-body density operator (dots), and with the inclusion of  MECs (triangles).}
\label{fig:5}       
\end{figure}

We can first look at the 3NF effects on the CSR. In Table \ref{tab:1}, we show the
CSR for different momentum transfers obtained with the AV18 and the AV18+UIX potential.
Clearly, the UIX force leads to a quenching of the CSR, especially at low momentum transfer, consistent with what was observed in the response function. At $q=50$ MeV/c the quenching on the overall strength is of about 20$\%$.

We can now  concentrate on $f_{pp} (q)$, since it contains the interesting
physical information about the proton-proton correlation function and since we can compare it with previous calculations. 
In Fig.~\ref{fig:5}, $f_{pp} (q)$ is shown in comparison with the results of Ref.~\cite{SICK} obtained 
with the same potential.
In Fig.~\ref{fig:5}, it is interesting to observe the perfect agreement of the two results at low $q$, where the one-body charge is used. 
We can also see that the contributions of two-body terms, called  meson 
exchange currents (MEC), to the charge density operator become negligible below $q= 300$ MeV.

Besides  the non relativistic expression ${\rho}({\bf q})\equiv {\rho}_{NR}({\bf q})$ of Eq.~(\ref{rho}), the first contribution due to relativistic effects lead to a one-body operator  given by 
\begin{eqnarray}
\label{rho_rel}
{\rho}_{RC}({\bf q})&=& -\frac{q^2}{8m^2}\sum_k \, e_k \exp{[i {\bf q} \cdot {\bf r}_k]}+ \\
\nonumber
&-&
i \sum_k \, \frac{e_k-\mu_k}{4m^2} {\boldsymbol \sigma}_k \cdot ({\bf q} \times {\bf p}_k)\exp{[i {\bf q} \cdot {\bf r}_k]} \,, 
\end{eqnarray}
where 
\begin{equation}
\label{muk}
\mu_k=\frac{1+\tau^3_k}{2}G^p_M(Q^2)+ \frac{1-\tau^3_k}{2}G^n_M(Q^2) \,,
\end{equation}
with $G_M^{p/n}$ being the proton and neutron magnetic form factors.
Here ${\boldsymbol \sigma}_k$ and ${\bf p}_k$ are the spin and the momentum 
of the $k-$th particle, respectively, and $m$ is the mass of the nucleon. The first term in Eq.(\ref{rho_rel}) is denominated by the Darwin-Foldy term, while the second is a spin-orbit one-body term.
In all previous Figures for $^4$He we have neglected relativistic contributions (RC).

\begin{figure}[!htb]
\centering
\includegraphics[width=1.0\columnwidth,angle=0]{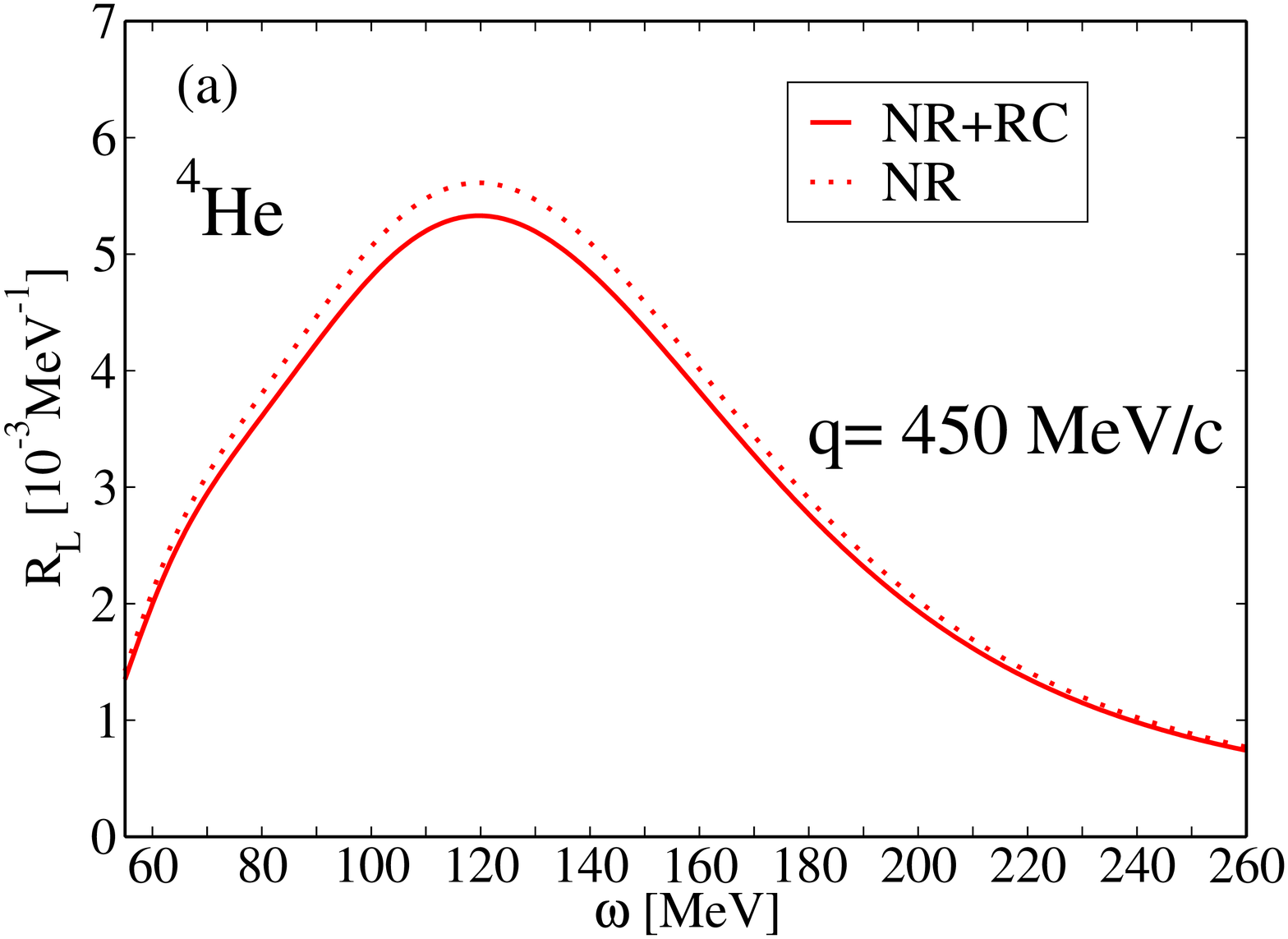}\\
~\\
\includegraphics[width=1.0\columnwidth,angle=0]{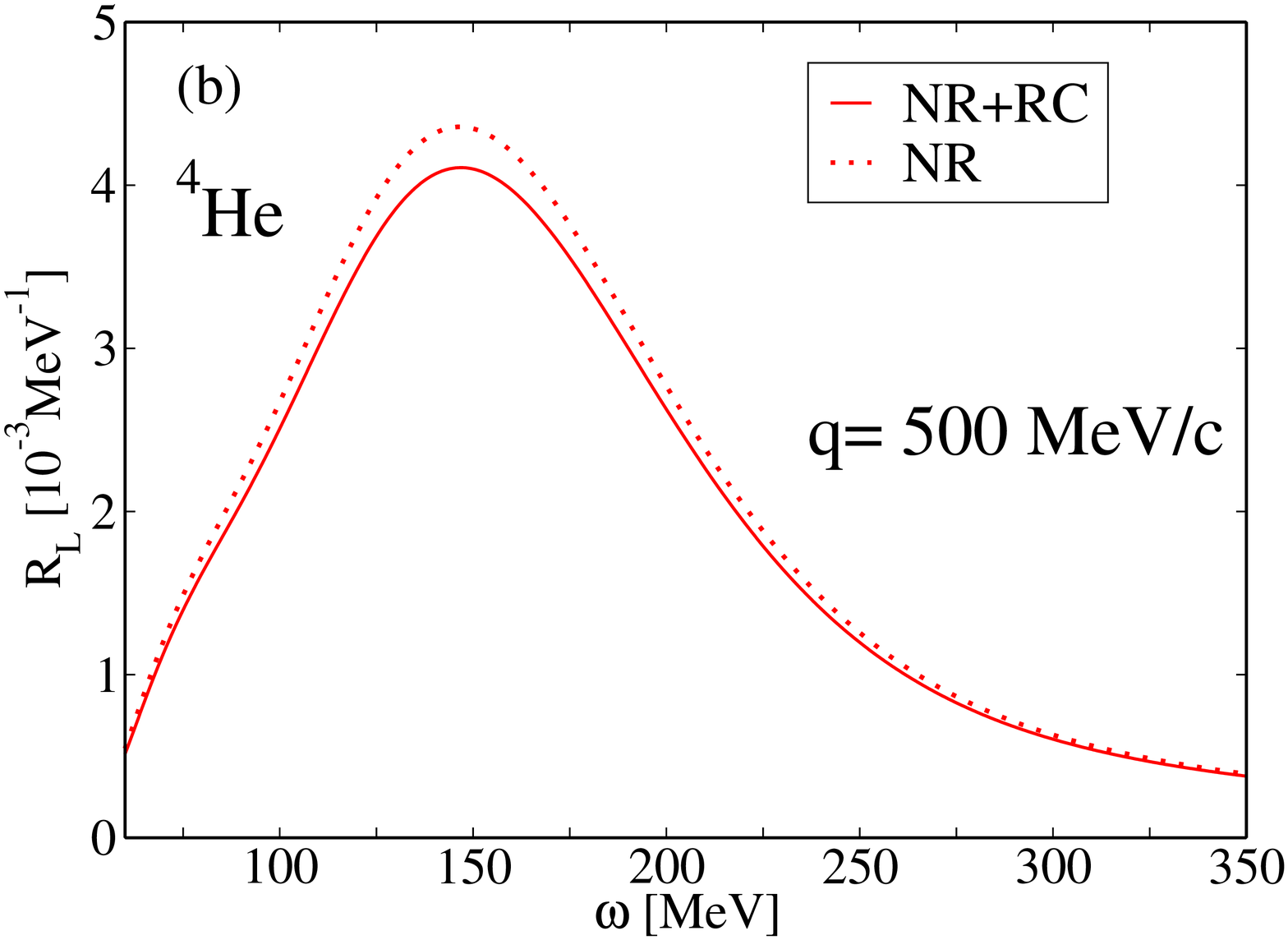}
\caption{Longitudinal response of $^4$He with the non relativistic (NR) one-body current of Eq.~(\ref{rho}) and with the Darwin-Foldy relativistic correction (RC) (see text), for $q=450$ and $500$ MeV/c. }
\label{fig:6}       
\end{figure}

While the inclusion of two-body currents in a realistic calculation of $R_L$ requires some labor,
 it is easier to include first order relativistic corrections.
We have included the Darwin-Foldy one body relativistic correction in Fig.\ref{fig:6}, where
its effect at $q=450$ and 500 is shown. It leads to a decrease of the cross section in the quasi-elastic peak by $5$ or $6 \%$.
The spin-orbit correction was found to be small in the case of $A=3$ \cite{RLA3}, and is probably small in $^4$He as well. Other relativistic effects, like wave function boosts have been found to be important in Ref.~\cite{RLA3}. We plan to investigate their effect on $^4$He in the near future.  

\subsubsection{The Transverse Response Function}

\begin{figure}[!h]
\centering
\includegraphics[width=1.0\columnwidth,angle=0]{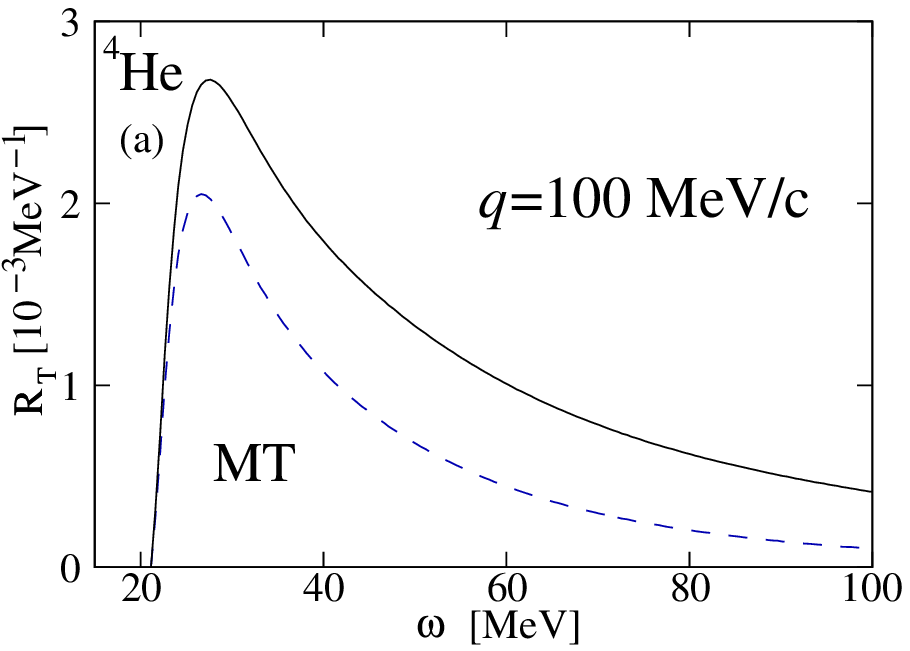}\\
\includegraphics[width=1.0\columnwidth,angle=0]{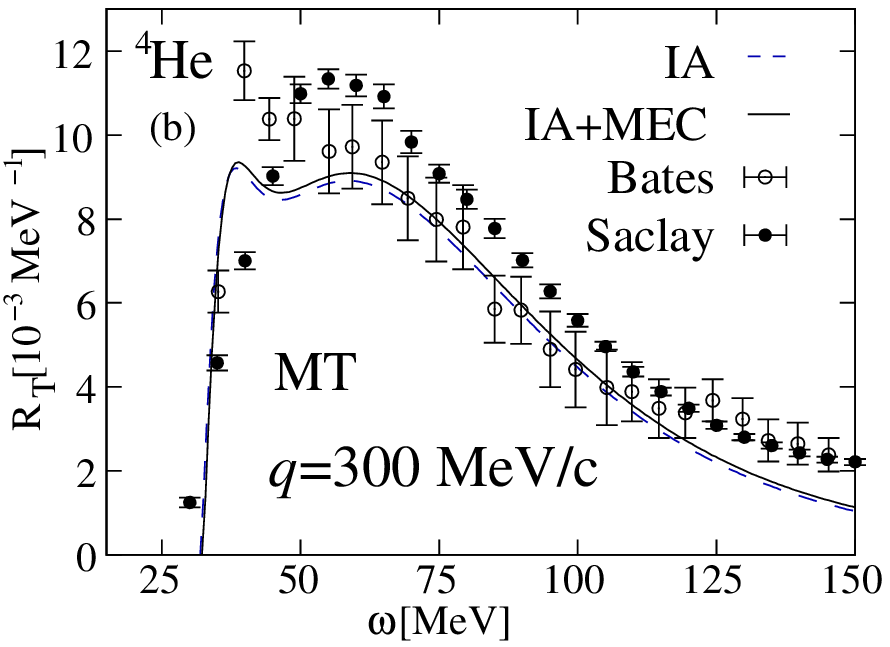}
\caption{Transverse response function $R_T$ of $^4$He at $q=100$ and $q=300$ MeV/c calculated with the MT potential: one-body current (dashed curve) response and one- plus two-body current response (solid curve) in comparison with experimental data from Bates \cite{Bates} and Saclay \cite{Saclay}.}
\label{fig_RT_He4}       
\end{figure}

The investigation of the transverse response function in the electron scattering reaction
is more involved, as it requires the knowledge of the transverse current operator.
The transverse response function is  defined as
\begin{equation}
R_T(\omega,{\bf q })=\int \!\!\!\!\!\!
\sum_f\left| \left\langle \Psi _{f}\right|{\bf j}_T({\bf q}) 
\left| \Psi _{0}\right\rangle \right| ^{2}\delta
\left(E_{f}+\frac{{\bf q}^2}{2M}-E_{0}-\omega \right)\,.
\label{rt}
\end{equation}
The transverse electromagnetic current density of a nucleus 
 can be
decomposed in a superposition of one- and many-body operators
\begin{equation}
{\bf j}_T({\bf q})= {\bf j}_{(1)}({\bf q})+{\bf j}_{(2)}({\bf
q})+ \dots \,.
\end{equation}
The non-relativistic one-body
 current density consists of convection
and spin currents 
\beq
{\bf j}_{(1)}({\bf q})={\bf j}^c_{(1)}({\bf q})+{\bf j}^s_{(1)}({\bf q})\,,
\eeq
with
\beqa
\label{currents}
{\bf j}^c_{(1)}({\bf q})&=& \frac{1}{2m}\sum_k e_k \,
\{{\bf p}_k,   \exp{[i {\bf q} \cdot {\bf r}_k]}  \}\,,\\
\nonumber 
{\bf j}^s_{(1)}({\bf q})&=& i\frac{1}{2m} \sum_k 
\mu_k\,({\boldsymbol \sigma}_k \times {\bf
q}) \exp{[i {\bf q} \cdot {\bf r}_k]}\,, 
\eeqa
where $e_k$ and $\mu_k$ are the same as in Eq.~(\ref{ek}) and (\ref{muk}), respectively.
Even in a non-relativistic
approach, the two-body term ${\bf j}_{(2)}$ does not vanish. 
In fact, a momentum and/or isospin dependent two-body interaction
$V$ requires the existence of a two-body current  operator, in order to satisfy the continuity
equation, which in coordinate space, reads 
\beq
\boldsymbol{ \nabla } \cdot {\bf j}_{(2)}({\bf
x})=-i[V,\rho({\bf x})]\,, \label{continuity_eq}
\eeq 
where $\rho$ is the Fourier transform of the one-body charge in Eq.~(\ref{rho}).
As is well known, this relation is not sufficient to determine
uniquely the two-body current. Therefore one needs in principle a
dynamic model for the nuclear potential which reveals the underlying
interaction mechanism. Such a model is supplied, for example, by the meson exchange
picture of the NN force.
But even for a phenomenological potential a consistent MEC can be constructed by expanding the potential in  meson exchange-like terms.

Investigations of $R_T$ with the LIT  have been done for $A=4$ and $A=3$ nuclear targets.
For the $\alpha$-particle only calculations with semi-realistic forces have been done \cite{ee'MT} so far, using central potentials with spin-isospin dependence, like the Malfliet-Tjon (MT) \cite{MaT69}.
In Figure \ref{fig_RT_He4}, we show our main results for $q=100$ and $q=300$ MeV/c. 
We present a calculation where  only the one-body current operator is used, called the impulse approximation (IA), in comparison  to one where  consistent MECs are included. Quite strong effects of the MEC are found at low-momentum transfer, where at energies of about 100 MeV, the two-body currents
lead to an enhancement of the strength by a factor of 4 with respect to the IA.
We also observed that at this low momentum transfer the spin current in Eq.~(\ref{currents}) does not dominate strongly: it only leads to half of the total IA strength \cite{ee'MT}. That is why
$R_T$ at low $q$ shows sensitivity to the convection current in Eq.(\ref{currents}) as well as to the MEC.
 Unfortunately, to the best of our knowledge, no experimental data are available for these kinematics. At larger momentum transfer, where data are found in literature, only a small effect due to the MEC is found. Here we observe that the spin current strongly dominates the strength at these kinematics. Moreover, the weak MEC effect is possibly due to the semi-realistic nature of both the potential and the consistent MEC, which included only fictitious scalar meson exchanges and no $\pi$- and $\rho$-exchange currents. 

Realistic calculations with the LIT have been done, instead, 
on the $A=3$ body systems ($^3$H and $^3$He) \cite{Sara2008} and \cite{RT09}. 
The interest of this kind of study is not  only to investigate the role of 3NFs.
The electron scattering reaction on the  $^3$He nucleus, in fact, is of particular importance, since it can also serve as effective neutron target.

In \cite{Sara2008} the BonnA \cite{BonnRA}  two-body potential was chosen, due to its meson-exchange character. In this way, meson exchange currents could be determined relatively uniquely.  The Hamiltonian was also augmented by an additional three-nucleon force, the TM'. MEC of $\pi$- and $\rho$-exchange nature have been used. In a subsequent calculation \cite{RT09}, the AV18+UIX force model was used, instead. In this case consistent MEC for the AV18 potential were introduced using the Arenh\"{o}vel and Schwamb model \cite{nonno-schwamb}, where an expansion of the potential in terms of fictitious meson exchange is done. 
Theoretical results were obtained for different current operators. A one-body current as in Eq.~(\ref{currents}) was used with the addition of one-body relativistic terms \cite{RT09} and MECs.
\begin{figure*}[!htb]
\centering
\includegraphics[width=2.0\columnwidth,angle=0]{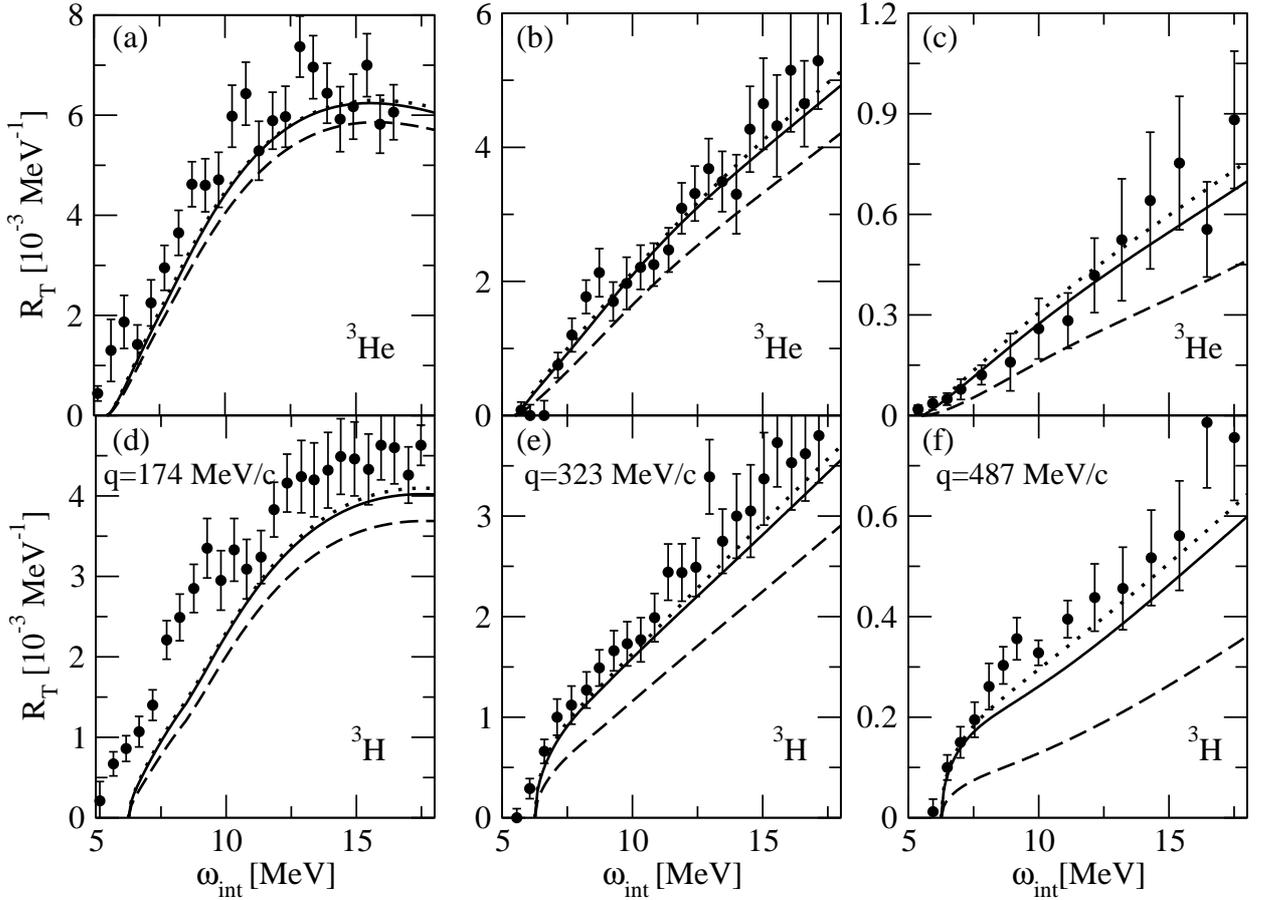}\\
\caption{Comparison of experimental \cite{Retzlaff} and theoretical results for the threshold $^3$He (upper panels) and $^3$H (lower panels) transverse response function $R_T$ with the AV18+UIX. Theoretical curves show different current operators: relativistic one-body current (dashed), relativistic one-body current + MEC (solid), non relativistic one-body current + MEC (dotted).
}
\label{fig9}       
\end{figure*}
In Fig.~\ref{fig9}, we compare theoretical results from the AV18+UIX potential with the experimental data from Retzlaff  {\it et al.}~\cite{Retzlaff}. Both the experimental data and the theoretical curves are plotted as a function of the intrinsic energy $w_{int}$, i.e. the recoil energy $\frac{{\bf q}^2}{2M}$ written in the formula (\ref{rt}) has not been added.
In the case of $^3$He one has a rather good agreement of theoretical and experimental results at the higher $q$ values. The MEC contribution is essential to reach agreement with the experiment and is pretty large at the threshold. At $q=174$ MeV/c the theoretical prediction for $R_T$ underestimates the data below 11 MeV. In the case of $^3$H the comparison of theory with respect to experiment looks worse, with a systematic  underestimation of data, which gets larger at the lowest $q$-value. Relativistic contributions are small, but visible already at $q=323$ MeV/c.
In \cite{RT09} it was also observed that different force models, like the BonnA+TM' and the AV18+UIX lead to very similar result.

\section{Weak Observables}
\label{sec:3}
In this section we deal with reactions induced by neutrino interactions on nucleonic matter.
Such processes are relevant to astrophysics, for example in the context of core-collapse supernovae.
The present theory of supernovae explosion still presents open questions about the explosion mechanism itself and also regarding  late nucleosynthesis processes. To better understand
these aspects, more microscopic calculations and improved simulations are needed.
Since, often no experiments are possible to simulate the conditions of a supernovae,   few-body calculations can provide valuable 
insight on many relevant reactions, that can  then be used as 
input 
for hydrodynamic simulations of the explosion. 
We will first discuss about inelastic neutrino reactions  of  $^4$He and secondly about
neutrino-antineutrino bremsstrahlung in neutron matter.

\subsection{Neutrino Scattering}

Core-collapse supernovae are believed to be neutrino driven explosions of massive stars.
At the end of its life a massive star ($M~>8~M_{\odot}$, where $M_{\odot}$ is the solar mass) presents a typical onion-like structure with layers containing ashes of
different burning stages, ending up with an iron core at the center. As the iron core grows and 
becomes gravitationally unstable, it collapses until nuclear forces halt the collapse and an 
outgoing shock wave is created. The most likely scenario is that the shock wave that propagates in the outer layers is stopped at some point due to energy loss via the dissociation of nuclei
into mainly free nucleons and $\alpha$-particles. One mechanism that can trigger a revival of the shock is given by interactions of neutrinos
coming out of the proto-neutron star (the remnant at the center of the iron core) with matter in the outer layers. In such a delayed explosion scenario, the neutrino energy transfer occurs mainly via elastic scattering of $\nu$s with matter (free nucleons and nuclei).
Nevertheless, other reactions can take place and contribute to the energy transfer.
One of them is the inelastic excitation of light nuclei, like $^2$H, $^3$He, $^3$H, and $^4$He
\cite{doron_achim}, \cite{almudena}.
The hot dilute gas above the proto-neutron star and the region below the accretion shock contains up to $70\%$ of $^4$He. That is why the inelastic excitation of $^4$He via neutrino scattering is
an important process to study in such a context. 

\begin{figure}[!h]
\centering
\includegraphics[width=0.7\columnwidth,angle=0]{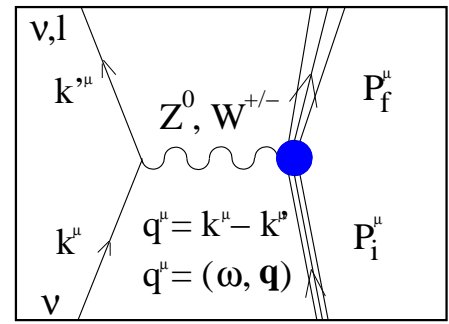}
\caption{Feynman diagram of the inelastic neutrino scattering off a nucleus,
where all neutrino flavors can be involved.}
\label{diag_neutr}       
\end{figure}

Neutrinos are produced  in flavor equilibrium inside the proto-neutron star. The characteristic temperatures of the emitted neutrinos are though different, due to the different reactions they undergo. They are about 6-10 MeV for $\nu_{\mu,\tau} (\bar{\nu}_{\mu,\tau})$, 5-8 MeV for $\bar{\nu}_e$ and 3-5 MeV for $\nu_e$. Thus, mainly  $\nu_{\mu,\tau}$ and $\bar{\nu}_{\mu,\tau}$ can carry  more than 20 MeV of energy, which is the amount needed to dissociate $^4$He via an inelastic neutral current interaction.
The $\nu_e$ and $\bar{\nu}_e$ can also contribute to a dissociation of the $\alpha$-particle via a charged current interaction, but in a less important way due to their lower energy. 
Even though a number of studies have been done
in the past to estimate these cross sections, the first ab-initio few-body study was done by Gazit and Barnea \cite{Doron2007} using the LIT and EIHH methods. 
A calculation was done for the energy dependent
{\it inclusive} inelastic cross section for $^4$He($\nu_x$, $\nu'_x$)$^4_2$X, $^4$He($\bar{\nu}_x$, $\bar{\nu}'_x$)$^4_2$X, $^4$He($\bar{\nu}_e$, $e^+$)$^4_1$X, and $^4$He($\nu_e$, $e^-$)$^4_3$X, where 
$x=e,\mu,\tau$ and $^A_ZX$ is the final state of $A$-nucleons with $Z$-protons.
The corresponding Feynman diagram is represented in Fig.~\ref{diag_neutr}.
\begin{figure}[!htb]
\centering
\includegraphics[width=1.0\columnwidth,angle=0]{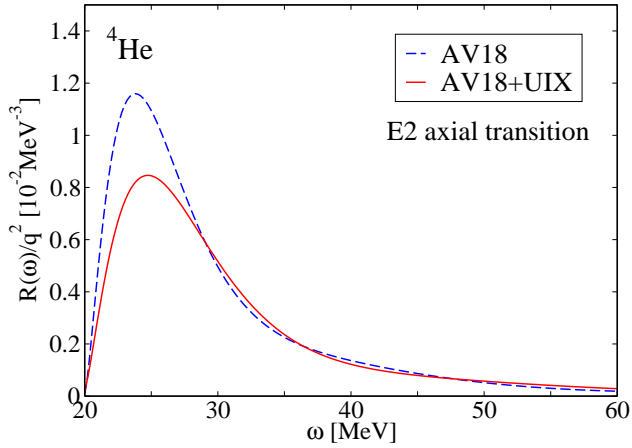}
\caption{Axial quadrupole transition  in the neutral neutrino scattering off $^4$He as a function of the excitation energy $\omega$. Calculations are performed with the AV18 and the  AV18+UIX potential and are taken from Ref.~\cite{doron_priv}. }
\label{fig:quad_nu}       
\end{figure}

After a multipole decomposition of the weak current, the cross section for the neutrino scattering process can be written as \cite{Doron2007}
\begin{eqnarray}
\label{neutr_cs}
\left(\frac{d\sigma^a}{dk_f}\right)_{\nu (\bar{\nu)}}&=&\frac{2G^2}{2J_i+1}k_f^2 F^a(Z_f,k_f)\\
\nonumber
&& \times \int d{\bs \epsilon}\int_0^{\pi}\sin{\theta}~\delta\left({\bs \epsilon -\omega + \frac{q^2}{2M}}\right)\\
\nonumber
&&\times 
\left\{ \sum_{J=0}^{\infty}\left[ X_{{C}}R_{{C}_J}+ X_{{L}}R_{{L}_J} -  X_{{C}{L}}
{\rm Re}R_{{C}_J^*{L}_J^*} \right] \right. \\
\nonumber
&&\times 
\left. \sum_{J=1}^{\infty}\left[ X_{{M}}R_{{M}_J}+ X_{{E}}R_{{E}_J} \mp  X_{{E}{M}}
{\rm Re}R_{{E}_J^*{M}_J^*} \right]\right\}\,, 
\end{eqnarray}
where $G$ is the Fermi weak coupling constant, $J_i$ is the initial angular momentum of the nucleus, $k_f$ is the momentum of the outgoing lepton, $Z_f$ is the charge of the residual system and $F$ is the Coulomb function. The superscript $a$ denotes the isospin component,
with $a=0$ for the neutral current and $a=\pm1$ for the charged current. $X_{{O}_1{O}_2}$ is 
the lepton kinematical function \cite{walechka} and $R_{{O}_1{O}_2}$ is the response function to the 
operators ${O}_1$ and ${O}_1$, where the notation $C_J,L_J,M_J$ and $E_J$ stands for Coulomb,
longitudinal, magnetic and electric multipoles of order $J$.

The calculation was done using the AV18 two-body force augmented by the UIX three nucleon
forces. The effect of the 3NF on astrophysically relevant observables
is an interesting one to investigate. 
In Figure~\ref{fig:quad_nu}, the response function of the dominant multipole (the axial  $E2$) in Eq.(\ref{neutr_cs})
for the neutral interaction cross section is shown as a function of the transferred energy $\omega$ for the AV18 and AV18+UIX potential. 
The numerical accuracy of these calculations is of the order of 1$\%$.
It is interesting to note that also in this case one observes a quenching effect due to 
the UIX force. This is analogous to what was found for the electron scattering observable, even though we are dealing with two different current operators.

The effect of MEC was also investigated using an hybrid approach: the weak current was constructed within EFT and then used with wave functions obtained from the traditional potentials AV18 and UIX. The axial current introduces a dependence of the MEC on a new low-energy constant,
denoted with $\hat{d}_r$, which was calibrated on the triton half life.
\begin{figure*}[!t]
\centering
\includegraphics[width=2.0\columnwidth,angle=0]{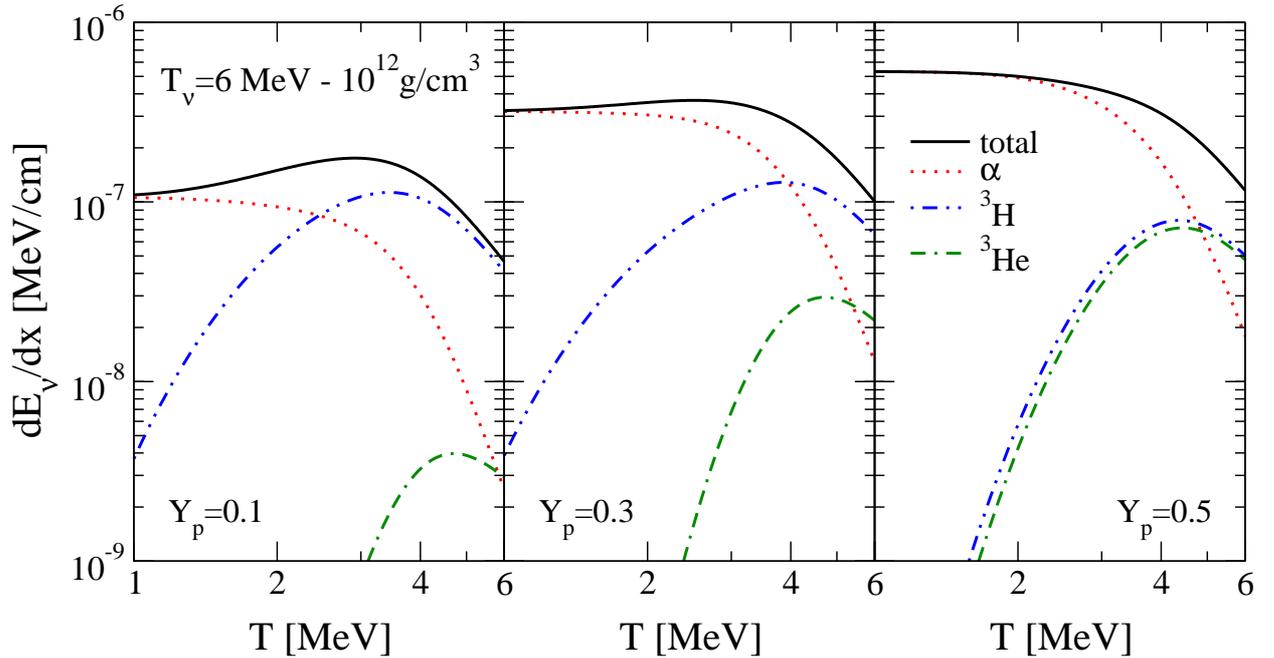}
\caption{Neutrino energy loss $dE_{\nu}/dx$ for inelastic excitations of $A=3,4$ nuclei as a function of the matter temperature $T$ and at density of $10^{12}$ g/cm$^3$. Neutrino energy spectra are assumed to be Fermi-Dirac distributions with temperature $T_{\nu}=6$ MeV. Different proton fractions are shown. Figure taken from Ref.~\cite{doron_achim}.}
\label{fig_achim}       
\end{figure*}
The  effect of the 3NF and the MEC on the temperature averaged neutral cross section
\begin{equation}
\langle \sigma^0_x \rangle_T=\frac{1}{2}\frac{1}{A}\langle \sigma^0_{\nu_x} + \sigma^0_{\bar{\nu}_x}\rangle_T
\end{equation}
 is shown in 
Table \ref{tab:2}.  Here the $\langle \dots \rangle_T$ symbol indicates the temperature average
\begin{equation}
\label{t_ave}
 \langle \sigma^0 \rangle_T = \int_{w_{th}}^{\infty} d\omega \int dk_i f(T_{\nu},k_i) \frac{d\sigma^0}{dk_f} \,,
\end{equation} 
which 
is obtained by integrating the cross section in energy and assuming that neutrino spectra are described by Fermi-Dirac distributions $f$ with temperature $T_{\nu}$. Here $k_i$ is the initial momentum of the neutrino.
One observes that the low-energy cross section is very sensitive to the nuclear force model,
while MECs are contributing in a negligible way. This is explained by the fact that the multipole
where MECs are expected to contribute the most is the Gamow-Teller, which is suppressed in $^4$He.

\begin{table}
\caption{Temperature averaged neutral current inclusive inelastic cross section per nucleon in
case of $^4$He as a function of the neutrino temperature. Data in [10$^{-42}$ cm$^2$] taken from \cite{Doron2007}.}
\label{tab:2}       
\begin{tabular}{cccc}
\hline\noalign{\smallskip}
T [MeV] & AV18 & AV18+UIX & AV18+UIX+MEC\\
\noalign{\smallskip}\hline\noalign{\smallskip}
4 & 2.31 10$^{-3}$ &1.63  10$^{-3}$ & 1.66  10$^{-3}$\\
6 & 4.30 10$^{-2}$ &3.17  10$^{-2}$ & 3.20  10$^{-2}$\\
8 & 2.52 10$^{-1}$ &1.91  10$^{-1}$ & 1.92  10$^{-1}$\\
10& 8.81 10$^{-1}$ &6.77  10$^{-1}$ & 6.82  10$^{-1}$\\
12& 2.29        &1.79         & 1.80\\
14& 4.53        &3.91         & 3.93\\
\noalign{\smallskip}\hline\noalign{\smallskip}
\end{tabular}
\end{table}

One could ask the question whether the inelastic excitation of nuclei other than the $\alpha$-particle can contribute to the neutrino energy transfer to the shock wave in supernovae. 
In order to tackle that question, one not only needs to know the inelastic cross section, but also the composition of the nucleonic matter.
The equations of state (EOS) typically used in supernovae simulations include neutrons, protons,
$\alpha$-particles and some representative heavy nuclei. In a recent study by O'Connor {\it et al.} \cite{doron_achim}
the virial EOS was extended to include also $^3$H and $^3$He nuclei and predict $A=3$ mass fractions near the neutrino sphere in supernovae. 
Neutrino cross sections of $A=3$ inelastic excitation were calculated with the LIT and EIHH, using the same potentials and MEC as done for the $\alpha$-particle.
An important fact was found: even though the $\alpha$-particles are more abundant,  the loosely bound $A=3$ nuclei dominate the energy transfer at low-density via inelastic reactions.
MECs were also found to be important: they contribute up to 16$\%$
in the thermal averaged cross section at temperature of 1 MeV. This is explained by the fact that the threshold cross section (which is dominant at low $T_{\nu}$) is largely influenced by the 
the axial dipole transition, where MECs are important.  

In Figure \ref{fig_achim}, the neutrino energy losses, 
\begin{equation}
\frac{dE_{\nu}}{dx}= n_b \sum_{i=^3{\rm H},^3{\rm He},^4{\rm He}} x_i \langle \omega \sigma^0 \rangle_{i,T_{\nu}}\,,
\end{equation}
are shown. Here $E_{\nu}$ is the neutrino energy, $n_b$ is the baryon density, $x_i=\frac{A_in_i}{n_b}$ is the mass fraction and  $\langle \omega \sigma \rangle_{i,T_{\nu}}$
is the energy weighted temperature averaged cross section (same as in Eq.~(\ref{t_ave}) with an energy weight).
The neutrino energy loss for inelastic excitations of $A=3,4$ nuclei was calculated as a function of the matter temperature $T\sim 4$ MeV and at density of about $10^{12}$ g/cm$^3$, which are typical conditions  near the neutrino-sphere. Neutrino energy spectra are assumed to be Fermi-Dirac distributions with temperature $T_{\nu}=6$ MeV. Different proton fractions $Y_p=0.1, 0.3, 0.5$ were investigated. One observes that $A=3$ nuclei contribute significantly to the neutrino energy loss due to inelastic excitations at $T\ge4$ MeV. As such, they should be included in supernovae simulations.

\subsection{Neutrino-Antineutrino Bremsstrahlung}

Neutrino processes involving two nucleons  play a key role
in the physics of core-collapse supernovae and neutron stars. 
In particular, neutrino-pair bremsstrahlung and absorption, $N N 
\leftrightarrow N N \nu \overline{\nu}$, are important processes for the production
of muon and tau neutrinos
~\cite{Raffelt}.
\begin{figure}[!h]
\centering
\includegraphics[width=0.7\columnwidth,angle=0]{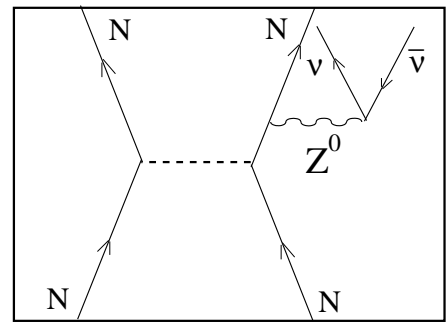}
\caption{Feynman diagram of the neutrino-antineutrino bremsstrahlung process on two neutrons. 
The interaction between the two nucleons is here represented by a one-pion exchange term
(pion denoted with the dashed line).}
\label{fig:brems_diag}       
\end{figure}
Due to the fact that neutrinos interact weakly, 
the rates for neutrino emission and absorption  are basically determined by the dynamic
response functions of strongly-interacting matter.
Supernova explosions are most sensitive to neutrino processes near the proto-neutron star, at subnuclear densities $\rho \sim \rho_0 /10$, where $\rho_0 = 2.8 \times 10^{14} $ g/cm$^3$ is the saturation density, and where matter is neutron rich.

The bremsstrahlung rate  was calculated  from NN phase shifts in Ref.~\cite{Hanhart}.
Here, we report on our recent calculation of the neutrino bremsstrahlung in pure neutron matter, where  we used chiral 
effective field theory (EFT) \cite{neutrino_paper}.
In Figure~\ref{fig:brems_diag}, a
Feynman diagram for  neutrino$-$ antineutrino bremsstrahlung is 
shown. Note that this is not the only  Feymnam diagram involved, as the $\nu$ $\bar{\nu}$ pair
can be produced from the ingoing left or right neutron, or from the outgoing left neutron, as well. 
If one considers all the four possibility, as done explicitly in \cite{FM}, one can observe that only
non-central contributions to strong interactions are relevant in such process.
Thus, in a chiral EFT language, tensor 
forces from pion exchanges and spin-orbit forces, are essential
for the two-nucleon response. 

In supernova and neutron star simulations, the 
standard rates for bremsstrahlung and absorption are based on
the one-pion exchange (OPE) approximation~\cite{FM,HR} for the interaction between the neutrons, which is 
represented by the dashed line in Fig.~\ref{fig:brems_diag}.
This is a reasonable starting point, since it represents the 
long-range part of nuclear forces, and it 
is the leading-order contribution in chiral EFT~\cite{Epelbaum}.
However, sub-leading non-central contributions are crucial for
reproducing NN scattering data~\cite{Epelbaum} and might be relevant
in neutrino bremsstrahlung as well.
We  systematically go beyond the OPE approximation and include 
contributions up to next-to-next-to-next-to-leading order 
(N$^3$LO) in chiral EFT. We find that shorter-range non-central
forces significantly reduce the neutrino rates for all relevant
densities.

We follow the approach to neutrino processes in nucleon matter
developed in Ref.~\cite{LPS}, which is based on Landau's theory
of Fermi liquids. This is a good approximation since the energies $\omega$
and momenta ${\bf q}$ transferred to the neutrinos are small in comparison to
the momenta of the neutrons. In addition, we consider degenerate conditions, where 
the temperature is small with respect to the Fermi energy. 
 The rate for bremsstrahlung for the dominant axial current process is given by~\cite{Raffelt}
\begin{equation}
\Gamma_{N N \leftrightarrow N N \nu \overline{\nu}}
= 2\pi \, n \, G^2 \, C_{\rm A}^2 \, (3-\cos\theta) \, 
R_{\rm A}(\om,\qq) \,,
\end{equation}
where $n$ denotes the neutron number density, $G$ the Fermi
coupling constant, $C_{\rm A} = - g_a/2 = - 1.26/2$ the axial-vector
coupling for neutrons, and $\theta$ here is the angle between the neutrino
and antineutrino momenta. The spin or axial response  $R_{\rm A}$ is 
given by~\cite{IP}
\begin{equation}
R_{\rm A}(\om,\qq) = \frac{1}{\pi n} \, \frac{1}{1-e^{-\om/T}} \,
{\rm Im} \, \chi_\sigma(\om,{\bf q}) \,,
\label{structspin}
\end{equation}
where $\chi_\sigma$ is the spin-density--spin-density response 
function, and we use units with $\hbar=c=k_{\rm B} = 1$.
We use a 
relaxation time approximation to solve the transport equation and 
obtain the response 
function~\cite{LPS}. 
Under this assumption, the bremsstrahlung neutrino rates 
$\Gamma_{N N \leftrightarrow N N \nu \overline{\nu}}$
depend on
$C_{\sigma}$, a quantity connected to the 
spin relaxation rate $\tau_\sigma$  
by~\cite{LPS,LOP}
\begin{equation}
\frac{1}{\tau_\sigma} = C_\sigma \, \bigl[
T^2 + (\om/2\pi)^2 \bigr]\,, 
\end{equation}
where
\begin{eqnarray}
C_\sigma& =& \frac{\pi^3 m^*}{6 \kf^2}  \biggl\langle 
\frac{1}{12} \, \sum\limits_{k=1,2,3} \\
\nonumber
&&{\rm Tr} \biggl[ \, {\cal A}_{{\bm \sigma}_1,{\bm \sigma}_2}(\kk,\kk') \,
{\bm \sigma}_1^k \bigl[ ({\bm \sigma}_1 + {\bm \sigma}_2)^k \, , \,
{\cal A}_{{\bm \sigma}_1,{\bm \sigma}_2}(-\kk,\kk') \bigr] \,
\biggr] \biggr\rangle \,.
\label{strace}
\end{eqnarray}
Here $m^*$ is the effective mass and $\kf$ the Fermi momentum. Also, ${\cal A}_{{\bm \sigma}_1,{\bm \sigma}_2}(\kk,\kk')$ is the
quasiparticle scattering amplitude multiplied by $N(0)=\frac{m^* \kf}{\pi^2}$,
the number of states at the Fermi surface,
$\kk = \pp_1 - \pp_3$ and $\kk' = \pp_1 - \pp_4$ are the nucleon
momentum transfers, and the average $\langle \ldots \rangle$ is over
the Fermi surface (for details see~\cite{LPS}). The commutator
with the two-body spin operator demonstrates that only non-central
interactions contribute.
The neutrino-pair bremsstrahlung and absorption
rates $\Gamma_{N N \leftrightarrow N N \nu \overline{\nu}}$
 are proportional to
$C_\sigma$ to a good approximation when $|\omega| \gg 1/\tau_\sigma$.
\begin{figure*}[!htb]
\centering
\includegraphics[width=1.2\columnwidth,angle=0]{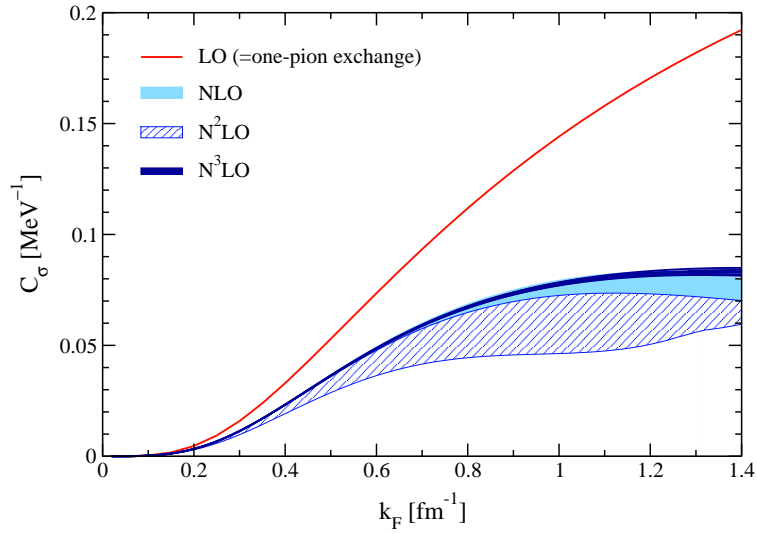}
\caption{ Spin relaxation rate given by $C_\sigma$
 as a function of Fermi momentum $\kf$ obtained
from chiral EFT interactions of successively higher orders~\cite{EGM}.
All results are for $m^*=m$.}
\label{orders}       
\end{figure*}
To calculate $C_\sigma$ we treat the strong interaction using the Born approximation and evaluate
the spin trace in the two-body spin space $| s \, m_s
\rangle$ using a partial-wave expansion. For $s=0$ the spin trace 
vanishes, thus only $s=1$ states contribute. We derived and expression
 (see Eq.~(6) in Ref.~\cite{neutrino_paper}) 
for $C_\sigma$ in terms of 
partial-wave matrix elements of the
strong interaction
 $\langle p |V_{\ell' 
\ell}^{j s}| p \rangle$, which depend on 
the relative momenta $p$.
For all results shown next, we found excellent
convergence when using 7 partial waves. In 
addition, we checked that the spin relaxation rate for OPE 
expanded in partial waves reproduces the known analytical result
(see Ref.~\cite{LPS}).

\begin{figure*}[!htb]
\centering
\includegraphics[width=1.2\columnwidth,angle=0]{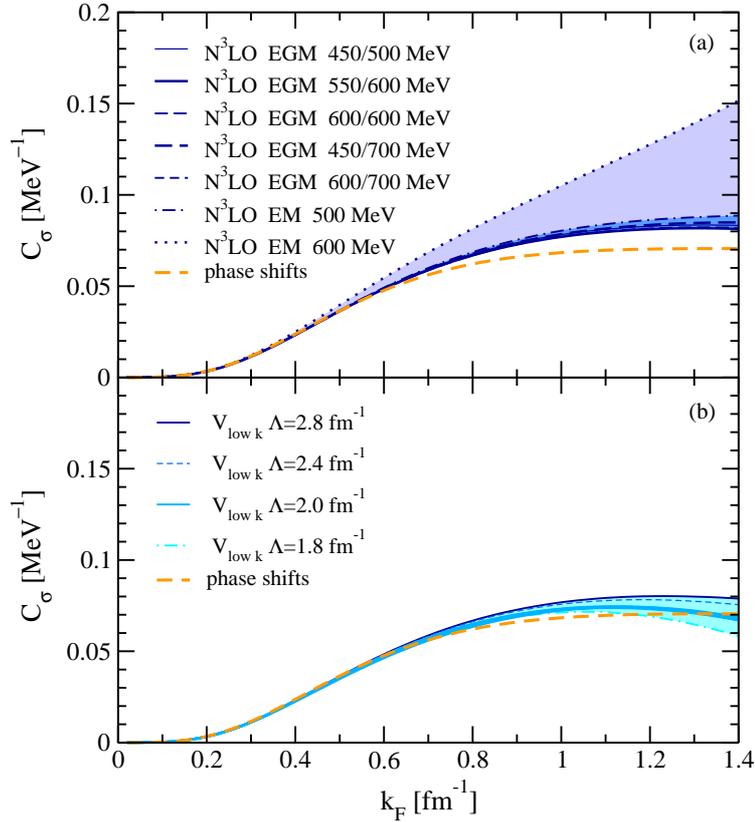}
\caption{The spin relaxation rate given by $C_\sigma$
 as a function of Fermi momentum $\kf$ obtained
from different chiral N$^3$LO potentials (see text) in comparison with
the spin relaxation rate obtained from empirical phase shifts (upper panel). 
The same observable calculated with
RG-evolved low-momentum interactions $\vlowk$ obtained from the EM 500 MeV chiral potential is shown in the lower panel for varying cutoff $\Lambda=1.8-2.8$ fm${^-1}$. The thickness of the curve for $\Lambda = 2.0$ 
fm$^{-1}$ is obtained evolving the 
seven different chiral N$^3$LO shown in the upper panel.
The effective mass here is $m^*=m_n=939.656$ MeV.
}
\label{cutoffs}       
\end{figure*}

In Fig.~\ref{orders}, we show the spin relaxation
rate calculated from chiral EFT interactions up to N$^3$LO as a function
of the Fermi momentum over a wide density range ($\kf 
= 1.4 \fmi$ corresponds to $\rho = m \kf^3/(3 \pi^2) = 1.6 \times 
10^{14} \gcmiq$). 
The leading order (LO) contribution includes 
OPE, which provides the standard
two-nucleon rates used in current supernova simulations.
The contact terms appearing at LO are purely central, thus do not contribute
to $C_\sigma$.
Our results
based on chiral EFT interactions of successively higher orders 
from Epelbaum {\it et al.}~(EGM)~\cite{EGM} show that OPE significantly 
overestimates $C_\sigma$ for all relevant densities. The bands at
next-to-leading order (NLO), N$^2$LO, and N$^3$LO provide an
estimate of the theoretical uncertainty, generated by varying
the cutoff $\Lambda$ as well as a spectral function
cutoff in the irreducible $2\pi$-exchange $\Lambda_{\rm 
SF}$~\cite{Epelbaum,EGM}. Chiral EFT interactions at N$^3$LO
accurately reproduce low-energy NN scattering~\cite{EGM,EM}, 
and at this order, $C_\sigma$ is practically independent of the 
N$^3$LO potential for these densities (see thin band at N$^3$LO in Fig.~\ref{orders}).
Most of the reduction of $C_\sigma$ occurs at the NLO level,
which includes the leading $2\pi$-exchange tensor force
and shorter range contact terms. We observe that at NLO the 
 $2\pi$-exchange added to the OPE enhances the $C_\sigma$, but this
is compensated by a further suppression due to the contact terms.

We have systematically investigated the sensitivity of this observable
on different chiral NN forces used in input.
In the upper panel of Fig.~\ref{cutoffs}, in addition to the band of the N$^3$LO potentials from \cite{EGM} of Fig.~\ref{orders}, we also show results for the Entem and Machleidt (EM)
N$^3$LO potentials with $\Lambda = 500$ and $600$ MeV~\cite{EM}. 
The EM $500$ MeV potential gives results which overlap with 
the EGM N$^3$LO band. The 
larger rate obtained with the EM $600$ MeV interaction is due 
to the failure of the Born approximation in this case.
We then  
also used the RG~\cite{Vlowk} to evolve 
N$^3$LO potentials to low-momentum interactions $\vlowk$ with 
$\Lambda=1.8$--$2.8 \fmi$, to extend the 
cutoff variation and estimate of the theoretical uncertainty.
The results are shown in the lower panel of Fig.~\ref{cutoffs}.
The RG evolution preserves the long-range pion 
exchanges and includes sub-leading contact
interactions, so that NN scattering data are reproduced. 
The thickness of the curve for $\Lambda = 2.0 \fmi$ is obtained 
after evolving all seven N$^3$LO potentials of the upper panel.
This shows the universality of $\vlowk$ and that, after 
high-momentum modes are integrated out, the particle-particle 
channel (for EM $600$ MeV) is rendered more perturbative~\cite{Vlowk}.

At low densities, two-nucleon collisions dominate and NN
scattering observables provide a model-independent result for the 
spin relaxation rate. To study this
we can replace the strong potential $V$  
 by the $T$ matrix, which is 
parameterized in terms of empirical phase shifts and mixing 
angles~\cite{BJ}, taken from the Nijmegen Partial Wave 
Analysis (nn-online.org).
For all densities in 
Fig.~\ref{cutoffs}, we find that the rates obtained in the Born 
approximation from chiral N$^3$LO potentials 
and $\vlowk$ interactions are close to
those from phase shifts. Combined with the RG insights, this
 demonstrates that the non-central part of
the strong neutron $-$ neutron amplitude is perturbative for 
these lower cutoff interactions.

The relatively small sensitivity  of $C_\sigma$ to the NN potential used in
input, allows us to fit a very simple analytical function
\begin{equation}
\frac{C_\sigma}{{\rm MeV}} = \frac{0.86 \, (\kf/{\rm fm}^{-1})^{3.6}}{1 
+ 10.9 \, (\kf/{\rm fm}^{-1})^{3.6}} \,,
\label{fit}
\end{equation}
that can be used in hydrodynamic simulations of supernovae explosions.
In fact, our main finding is that two-pion exchange interactions and 
shorter-range non-central forces, included beyond the LO in EFT, 
reduce the neutrino rates significantly, with respect to the standard
rates used in supernovae simulations. 
Future work will include neutron-proton mixtures and charged currents, to systematically
improve the neutrino physics input for astrophysics.
We also plan to investigate other many$-$body effects on collisions including 3NFs,
which are expected to be important at higher densities.

\section{Conclusions and Outlook}
\label{sec:4}

The investigation of electroweak reactions involving light nuclei has entered a new era, due to new theoretical advances and to the number of physical aspects one can study.
It allows one to tackle a variety of questions, ranging from more fundamental ones
on how nucleons interact with each other and  spanning to astrophysics, where one can 
learn about the role of nuclear  input in astrophysical objects.

Here, we have reported on the recent progress done 
in calculations of electroweak reactions on light nuclei, mainly using the LIT method in
conjunction with EIHH expansions of the wave function. In such way, bound and scattering observables are treated on an equal footing. The use of the LIT method is so far the only viable procedure 
to access information on the inelastic response of  $A>3$ nuclei for energies beyond the
three-body disintegration threshold. A direct calculation of all involved channels is in fact
impossible. The LIT method, by circumventing the problem via an integral transform and turning
it to the solution of a bound state equation, constitutes a very practical and yet effective
approach to obtain the exact response function for light nuclei.    

The excitation of nuclei with external electromagnetic probes is a fundamental tool
to get information about the nuclear dynamics via a comparison of theoretical predictions
with experiments.
Theoretical efforts  need to be addressed to the 
identification of many-body observables that are sensitive to 3NFs, that 
are not constrained by NN scattering.
We have shown that the photo-absorption reaction on $^4$He presents a relatively mild sensitivity to 3NF
in the energy range of the cross section peak.  Furthermore, the experimental situation on the $\alpha$-particle does not allow one to discriminate among difference force models.
On the other hand, the electron scattering reactions can potentially provide us with useful 
information  on the nuclear dynamics. Here in fact, we find a larger effect of  3NF, especially
at low-momentum transfer. In this region no experimental measurements are found in literature, but new data have been taken in Mainz and are to be analyzed \cite{Distler}. More work needs to be done from the theoretical point of view by using other potential models available and by performing further sensitivity studies. For example, calculations of $R_L$ and $R_T$ at low $q$ with
chiral EFT potentials are called for. If the new experimental data are precise enough, one could explore the possibility of calibrating low energy constants on such continuum observables. They are in fact expected to contain a richer information about nuclear dynamics than  simple  bound stable observables, like the binding energy, due to the fact that many channels are involved.
 
We have also shown recent results for weak observables where excitations of nucleonic matter 
are induced by neutrino interactions. Such processes are very important in astrophysical 
scenarios, like in supernovae explosions. 
We have reported on a recent ab-initio calculation \cite{Doron2007} of the $\nu$-$\alpha$ scattering with two and three-body forces, that 
can be used as input for supernovae simulations. 
One can observe that  such astrophysically relevant observables also display a dependence on three-body forces, which is interestingly  similar to that found in electron scattering.
We also have shown how a recent study \cite{doron_achim} pointed out the importance of $A=3$  disintegration as a way to contribute to the neutrino energy transfer to the shock wave in supernovae.
Finally, we have presented a study of the neutrino-bremsstrahlung process  using  chiral EFT potentials. We have shown how modern two-body Hamiltonians can improve our understanding of such
reaction rates.
Our results can be incorporated in simulations and are part of a larger project to systematically improve the nuclear microscopic input for supernovae simulations. 
Future work will address the role of many-body physics, including the effect of 3NFs, relevant at higher densities. 
Summarizing, since often no experiments are possible to simulate astrophysical conditions, all presented calculations show how  few-body physics can provide valuable input for hydrodynamic simulations of supernovae explosions.

\section{Acknowledgment} It is a pleasure to thank the organizers of the conference for a very interesting workshop and providing a stimulating environment.
I would like to thank Winfried Leidemann for providing the pictures about electron scattering off $A=3$ nuclei and Doron Gazit for making available the data on the inelastic neutrino cross section of $^4$He. I am thankful also to Achim Schwenk for useful conversations.
This work was supported in part by the
Natural Sciences and Engineering Research Council (NSERC) and by the
National Research Council of Canada.

\end{document}